\documentclass{aa}
\usepackage{natbib}         
\usepackage{graphicx}                    
\usepackage{amssymb}                    
\usepackage[usenames]{color}                         
\usepackage{txfonts}

\newcommand{\BE}{\begin{equation}}
\newcommand{\EE}{\end{equation}}
\newcommand{\BA}{\begin{eqnarray}}
\newcommand{\EA}{\end{eqnarray}}
 \newcommand{\fig}[1]{Fig.~\ref{fig_#1}}
 \newcommand{\figs}[2]{Figs.~\ref{fig_#1} and \ref{fig_#2}}
 
 \newcommand{\sect}[1]{Sect.~\ref{sec_#1}}
 
 \newcommand{\eq}[1]{Eq.~(\ref{eq_#1})}
 \newcommand{\eqs}[2]{Eqs.~(\ref{eq_#1}) and (\ref{eq_#2})}
 
\newcommand{\eg}{e.g.}
\newcommand{\ie}{i.e.}
\newcommand{\insitu}{in situ}

\newcommand{\degree}{\ensuremath{^\circ}}

\newcommand{\rmd}{{\rm d }}

\newcommand{\uvec}[1]{\hat{\bf #1}}

\newcommand{\bA}{\beta_{\rm A}}
\newcommand{\lA}{\lambda}
\newcommand{\iA}{i}  
\newcommand{\pA}{\phi}
\newcommand{\tA}{\theta}

\newcommand{\pl}{\mathcal{P}(\lambda)}
\newcommand{\pobs}{\mathcal{P}_{\rm obs}}
\newcommand{\pobsl}{\mathcal{P}_{\rm obs}(|\lA|)}
\newcommand{\prob}{\mathcal{P}}
\newcommand{\pvphi}{\mathcal{P}_{\varphi}}
\newcommand{\phimax}{\varphi_{\rm max}}

\newcommand{\ur}{\hat{\bf u}_{\rho}}
\newcommand{\up}{\hat{\bf u}_{\varphi}}

\begin{document}

\title{Global axis shape of magnetic clouds deduced from the distribution of their local axis orientation}

\titlerunning{Axis shape of magnetic clouds}

\author{M. Janvier\inst{1} \and P. D\'emoulin\inst{1} \and S. Dasso\inst{2,3}
            }
   \offprints{M. Janvier}
\institute{
$^{1}$ Observatoire de Paris, LESIA, UMR 8109 (CNRS), F-92195 Meudon Principal Cedex, France \email{Miho.Janvier@obspm.fr, Pascal.Demoulin@obspm.fr}\\
$^{2}$ Departamento de F\'\i sica, Facultad de Ciencias Exactas y Naturales, Universidad de Buenos Aires, 1428 Buenos Aires, Argentina \email{dasso@df.uba.ar}\\
$^{3}$ Instituto de Astronom\'\i a y F\'\i sica del Espacio, UBA-CONICET, CC. 67, Suc. 28, 1428 Buenos Aires, Argentina \\
}
   \date{Received ***; accepted ***}

   \abstract
   {Coronal mass ejections (CMEs) are routinely tracked with imagers
in the interplanetary space while magnetic clouds (MCs) properties are measured locally by spacecraft. However, both imager and \insitu\ data do not provide direct estimation on the global flux rope properties.}
   {The main aim of this study is to constrain the global shape of the flux rope axis from local measurements, and to compare the results from in-situ data with imager observations.     
   }
   {We perform a statistical analysis of the set of MCs observed by WIND spacecraft over 15 years in the vicinity of Earth.  We analyze the correlation between different MC parameters and study the statistical distributions of the angles defining the local axis orientation. 
With the hypothesis of having a sample of MCs with a uniform 
distribution of spacecraft crossing along their axis, 
we show that a mean axis shape can be derived from the distribution of the axis orientation. 
In complement, while heliospheric imagers 
do not typically observe MCs but only their sheath region, we analyze one event where the flux-rope axis can be estimated from the STEREO imagers.}
   {From the analysis of a set of theoretical models, we show that the distribution of the local axis orientation is strongly affected by the global axis shape. Next, we derive the mean axis shape from the integration of the observed orientation distribution.
This shape is robust as it is mostly determined from the global shape of the distribution.  
Moreover, we find no dependence on the flux-rope inclination on the ecliptic.  Finally, the derived shape is fully consistent with the one derived from heliospheric imager observations of the June 2008 event. 
   }
   {We have derived a mean shape of MC axis which only depends on one free parameter, the angular separation of the legs (as viewed from the Sun). This mean shape can be used in various contexts such as the study of high energy particles or space weather forecast.
   }


   \maketitle

\begin{figure*}[t!]    
\centering
\includegraphics[width=4.9cm, clip=]{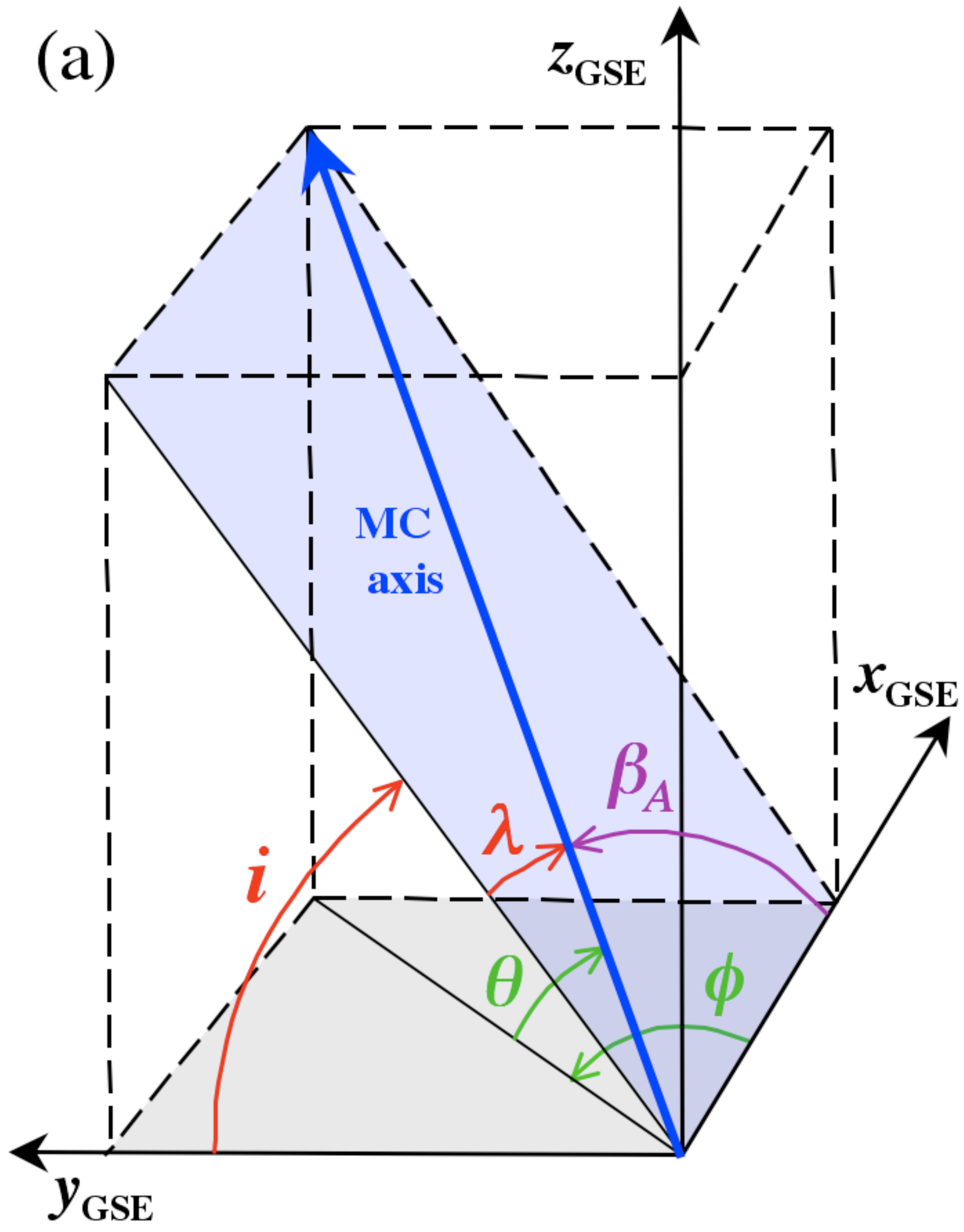}
\includegraphics[width=4.2cm, clip=]{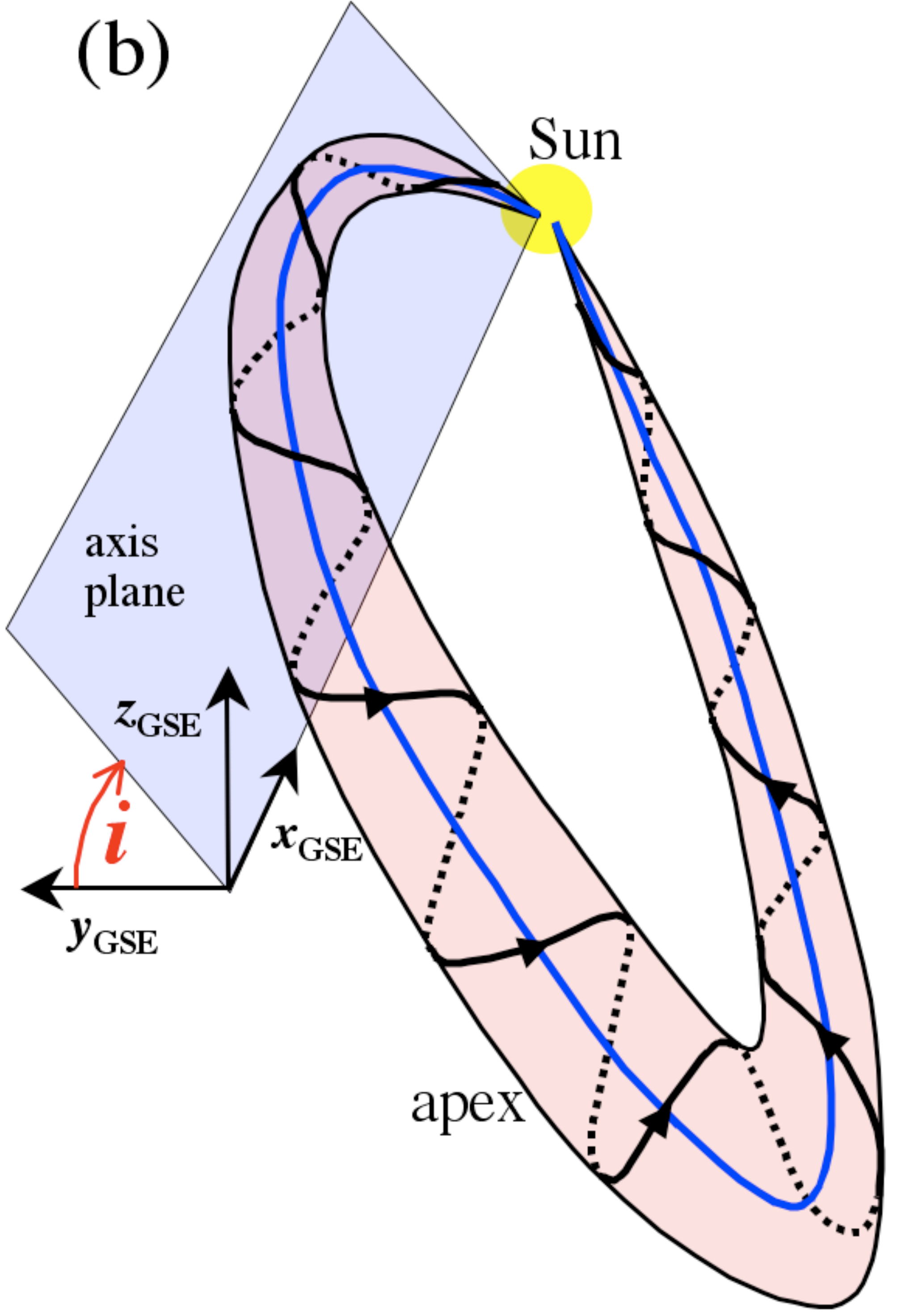}
\includegraphics[width=8.3cm, clip=]{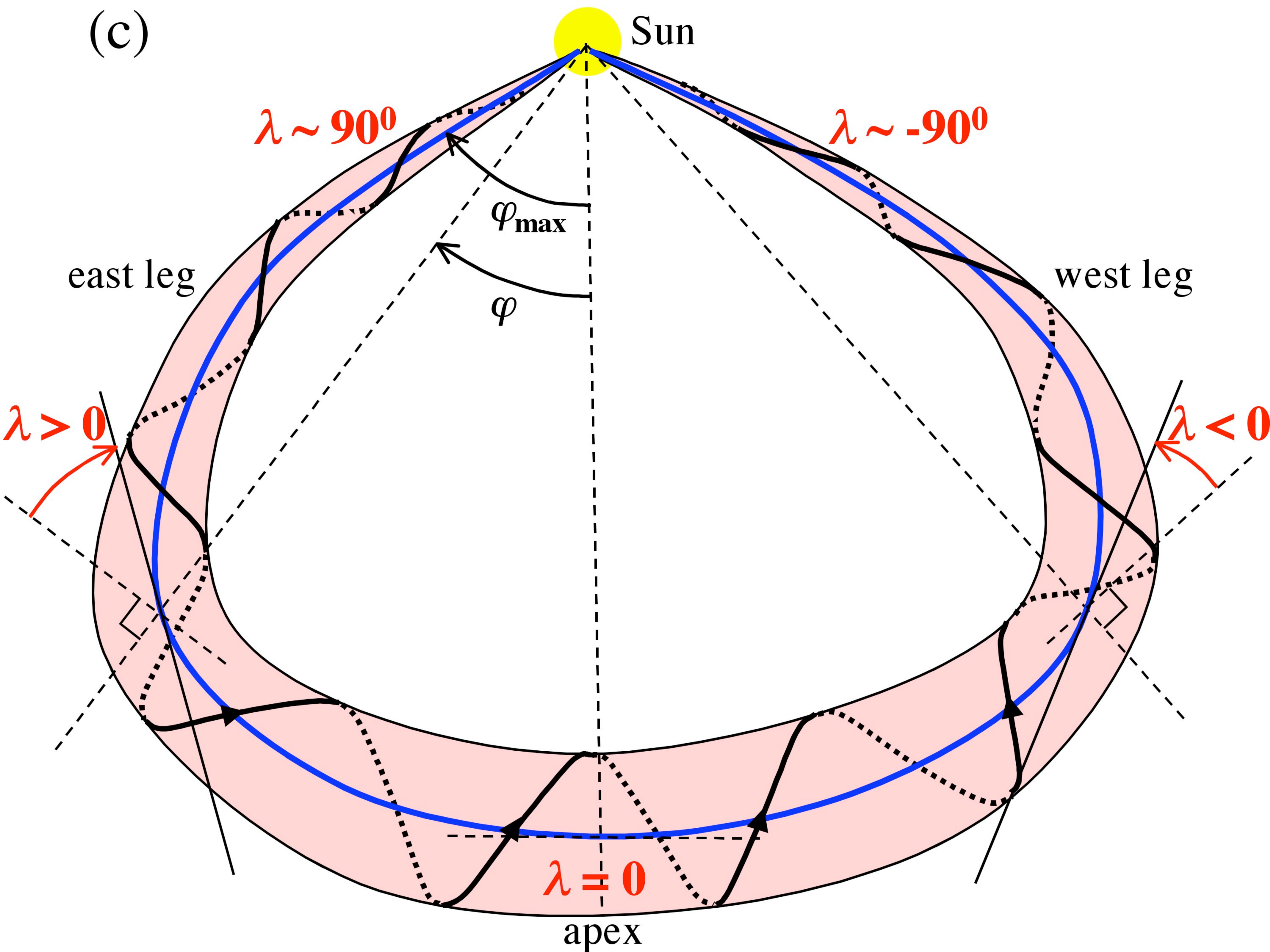}
\caption{Definitions of the angles and flux-rope geometry.
 {\bf a)} Schema defining the angles of the local MC axis direction. It is a local view of panel b. The unit vector $\uvec{x}_{GSE}$ points toward the Sun and $\uvec{z}_{GSE}$ is orthogonal to the ecliptic and northward. 
$\pA$ and $\tA$ are respectively the longitude and the latitude of the MC axis (spherical coordinates with the polar axis $z_{\rm GSE}$).  The axis direction can also be defined by $\iA$ and $\lA$ angles which are respectively the inclination and the position angle (spherical coordinates with the polar axis $x_{\rm GSE}$). $\beta_A$ is the cone angle defined from $\hat{x}_{GSE}$ to the MC axis. All the angles $\phi, \theta, \lambda$ and $i$ are drawn with positive values.
 {\bf b)} Schema showing the large scale meaning of $\iA$ when the flux-rope axis is in a plane (light blue, drawn northward of the radial Sun-spacecraft direction).
 {\bf c)} Schema showing the large scale meaning of $\lA$ and drawn in the plane of the flux-rope axis.  This plane in 3D is inclined by an angle $\iA$ on the ecliptic (left panels).  Examples of spacecraft trajectories across the flux-rope are shown with radial dashed lines supposing that the flux-rope is expanding radially away from the Sun. 
}
 \label{fig_schema}
\end{figure*}
 
\section{Introduction} 
\label{sec_Introduction}
 
Coronal magnetic configurations are frequently unstable and 
lead to coronal mass ejections (CMEs) propagating into the interplaneraty space \citep[see the reviews of][]{Pick06,Kleimann12}.   
Evidences of the presence of a twisted flux tube, or flux-rope, have been reported before the launch and especially during the time when the CME takes off \citep{Canou09,Guo10,Cheng11,Cheng13,Patsourakos13}. Then, MHD models of CMEs commonly include a flux-rope \citep[\eg][and references therein]{Forbes06,Aulanier12,Schmieder12}.  
Coronagraph observations visualize the denser regions of CMEs through the Thomson scattering of white-light by free electrons \citep[see][for reviews]{Howard11,Thernisien11}.  These observations are compatible with a flux-rope topology with the observed appearance depending on the relative orientation of the flux-rope with the line of sight \citep[\eg][]{Cremades04}.  
An approach was developed with a forward model having a dense shell around a flux-rope like shape and fitted visually to coronagraph images of CMEs \citep[][and references therein]{Thernisien06,Krall07,Thernisien11b}.  The method was developed to incorporate the two views from 
the Solar Terrestrial Relations Observatory (STEREO) spacecraft \citep{Wood09,Wood11}.   These advances were also supported by important developments of MHD simulations of flux-rope propagation as reviewed by \citet{Lugaz11}. 

Magnetic clouds (MCs) are detected within a fraction of interplanetary CMEs \citep[][and references therein]{Wimmer-Schweingruber06}.  Their main characteristic is a large and smooth rotation of the magnetic field direction.  This signature is classically interpreted by the presence of a twisted magnetic flux tube, simply called a flux-rope \citep[\eg][and references therein]{Lepping90,Burlaga95}.  However, the \insitu\ observations alone are not sufficient to firmly conclude that a flux-rope configuration is the unique possibility \citep{Al-Haddad11}, but combining \insitu, coronagraphic observations and forward modeling \citep{Krall07} enforce the presence of a flux-rope in all CMEs \citep{Xie13}.

The magnetic field and plasma measurements are only available along the spacecraft trajectory during the MC crossing. Then, various magnetic models can be proposed to gain information on the flux-rope cross section. Their free parameters are determined by a least square fit to the magnetic data obtained from the \insitu\ observations.  Such models can then provide the magnetic field distribution within the cross section as well as the local axis orientation of the flux-rope. The simplest and most used model is the cylindrical linear force-free field model also referred to as the Lundquist's model \citep[see \eg][]{Goldstein83,
Lepping90,Leitner07}. Extensions to non circular cross-section \citep[\eg][]{Vandas03,Demoulin09b}, or non force-free models \citep[\eg][]{Mulligan99b,Mulligan01,Hidalgo11} have been proposed without, so far, a model emerging as a standard for MCs. An alternative is to solve the magneto-hydrostatic equations in the MC frame with the magnetic data as boundary conditions for the integration procedure and with the hypothesis of local invariance along the axis. In such a model, the theoretical constraint that the plasma and axial field pressure should only depend on the magnetic flux function in the cross section is used to determine the local axis direction \citep[\eg ][]{Hu02,Sonnerup06,Isavnin11}.  
To summarize, all these approaches provide a magnetic model of the flux-rope cross section with a local invariance along the axis.
    
An extension of the approaches presented above was proposed with several models developed to incorporate the curvature of the flux-rope axis with a toroidal geometry \citep[keeping an invariance along the axis, \eg][]{Marubashi97,Romashets03,Marubashi07,Romashets09}. This is especially needed when the angle between the spacecraft trajectory and the local axis direction is small \citep[\eg][]{Marubashi12,Owens12}.  The inclusion of the toroidal geometry implies a larger number of free parameters for the model, and it is not yet demonstrated {how well} the data from a single spacecraft can constrain all of them, in particular the local curvature of the axis {that is important to obtain the global axis shape of the flux-rope axis}. {Such approach would benefit from well separated spacecraft as data from only two spacecraft provide more constrains to the toroidal model \citep{Nakagawa10}.}

{However, although} multi-spacecraft observations can provide a more 
complete set of data to analyze a flux rope configuration, there is 
only a very limited number of MCs that have been sampled along the flux-rope by at least two spacecraft \citep[see][for a review]{Kilpua11}.  
When the two spacecraft are separated with a significant angle (several 10\degree\ as seen from the Sun), the data can provide a rough estimation of the extension of the flux-rope \citep{Mulligan01,Reisenfeld03}. Then, tighter constraints require more spacecraft, but only very few MCs have been observed by at least three spacecraft crossing the flux-rope sufficiently close to its axis.  {When possible,} such cases allow the local determination of its axis orientation at distant regions along the axis if the spacecraft are sufficiently separated \citep{Farrugia11,Ruffenach12}.  Still, it remains unclear whether the different methods used to determine the axis orientation (see above) have large scatter/bias or if the flux-rope axis has a more complex shape than typically proposed \citep[compare \fig{schema}c to Fig.~12 of][]{Farrugia11}.  Finally, the case studied by \citet{Burlaga81}, with a MC scanned by four spacecraft, remains an exceptional case from which the flux-rope shape was constrained (see their Fig.~5). {The occurrence of correlated observations therefore remains too scarce to derive from such studies mean global properties of the flux rope in MCs.}

On the other hand, several types of study require a more global view of the flux-rope structure.  An example is the understanding of the crucial role played by the field line length in the time delay observed between particles of different energies during the propagation of high energy particles within MCs \citep[\eg ][]{Larson97,Masson12}.
Another example is the study aiming at relating the flux-rope properties to the 3D configuration of its solar source in a more complete way than with timing, orientation, and magnetic flux \citep[\eg\ as realized in][and references therein]{Nakwacki11}.
{So far,} simplified methods have been developed to get estimations of some global MHD quantities contained in MCs, such as magnetic helicity \citep{Dasso03,Dasso06,Dasso09b} or magnetic energy \citep{Nakwacki11}. {They modeled} the local {flux} tube of the cloud given from in situ observations. However, {a proper model of the global magnetic cloud shape will help improving their quantification}.

The determination of the 3D shape of a MC would need many spacecraft to sample it at as many locations as possible.  Since this is prohibitory costly, could we rather combine the information obtained on many MCs to derive a mean global configuration of MCs? Supposing a simply curved axis, the flux-rope axis direction provides an indication on the location where the flux-rope is crossed by the spacecraft (\fig{schema}c).   For example, an axis orthogonal to the radial direction (Sun-spacecraft) would mean that the flux rope is crossed at its apex (or nose), while a local axis more oriented in the radial direction would imply that the crossing is further away from the apex.  Then, observations of many MCs with various deduced local axis directions can sample flux ropes along their axis.

 In this study, we further analyze the above property to derive a mean axis shape for the set of studied MCs. This is done in three main steps. First, in \sect{obs}, we analyze the statistical properties of the set of MCs, testing the correlation between the MC parameters. We derive the statistical distributions of {the axis orientation parameters}, and test their robustness using various selection criteria on the MC parameters.  Second, in \sect{Simple}, we use an axis model to 
investigate the effect of the global axis shape on the distribution of the local axis orientation. Then, in \sect{Deduction}, we present the reverse procedure: we deduce the mean global axis shape from the observed distribution of the local axis orientation. 
  These results are complemented in \sect{Comparison} by our analysis of a well-observed event where the flux-rope extension and its axis can be constrained by heliospheric images and \insitu\ data. Here, we compare the axis shape deduced from the imager data with our results from the \insitu\ data of a MC set. Finally, in \sect{Conclusion}, we summarize our results and conclude on their implications.

\begin{figure*}
\centering
\sidecaption    
\includegraphics[width=12cm]{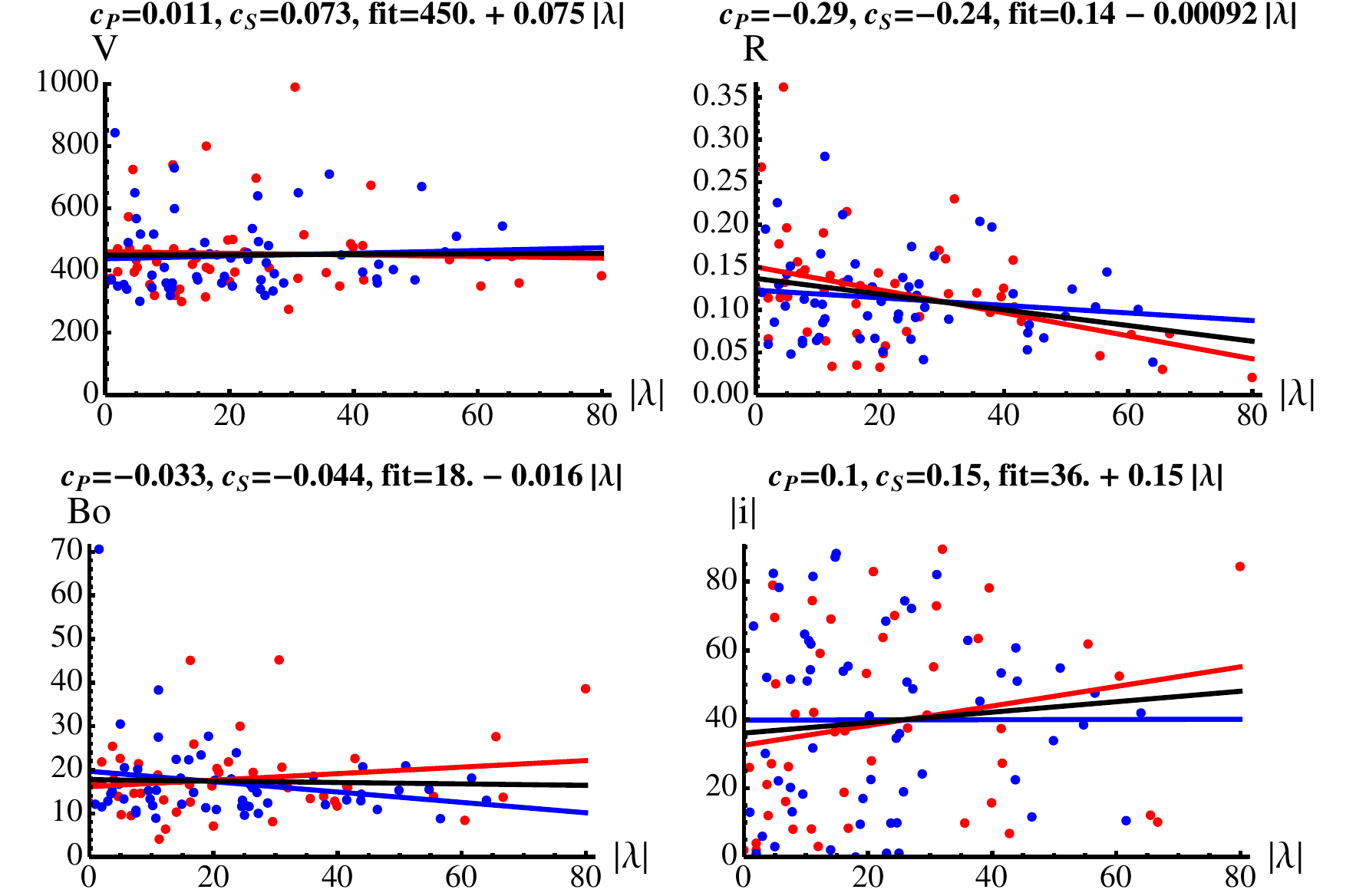}
 \caption{Properties of MCs observed at 1~AU versus the location angle ($\lA$ in degree). The correlations are shown for: the mean MC velocity ($V$ in km/s), the MC radius ($R$ in AU), the axial magnetic field strength ($B_0$ in nT), and the axis inclination ($\iA$ in degree) for the full set of MCs.  $\lA >0$ and $\lA <0$ are respectively shown in red and blue, and the abscissa,
$|\lA|$ allows to compare the two leg sides of the flux-rope (\fig{schema}c).  
The straight lines are linear fits to the data points (MCs) showing the global tendency. 
The results with the total MC set are shown in black (linear fit and top labels). $c_P$ and $c_S$ are respectively the Pearson and Spearman rank correlation coefficients, and ``fit'' is the least-square fit of a straight line (in black) to the full data set.}
\label{fig_corel_lambda}
\end{figure*}

\begin{figure*}
\centering
\sidecaption    
\includegraphics[width=12cm]{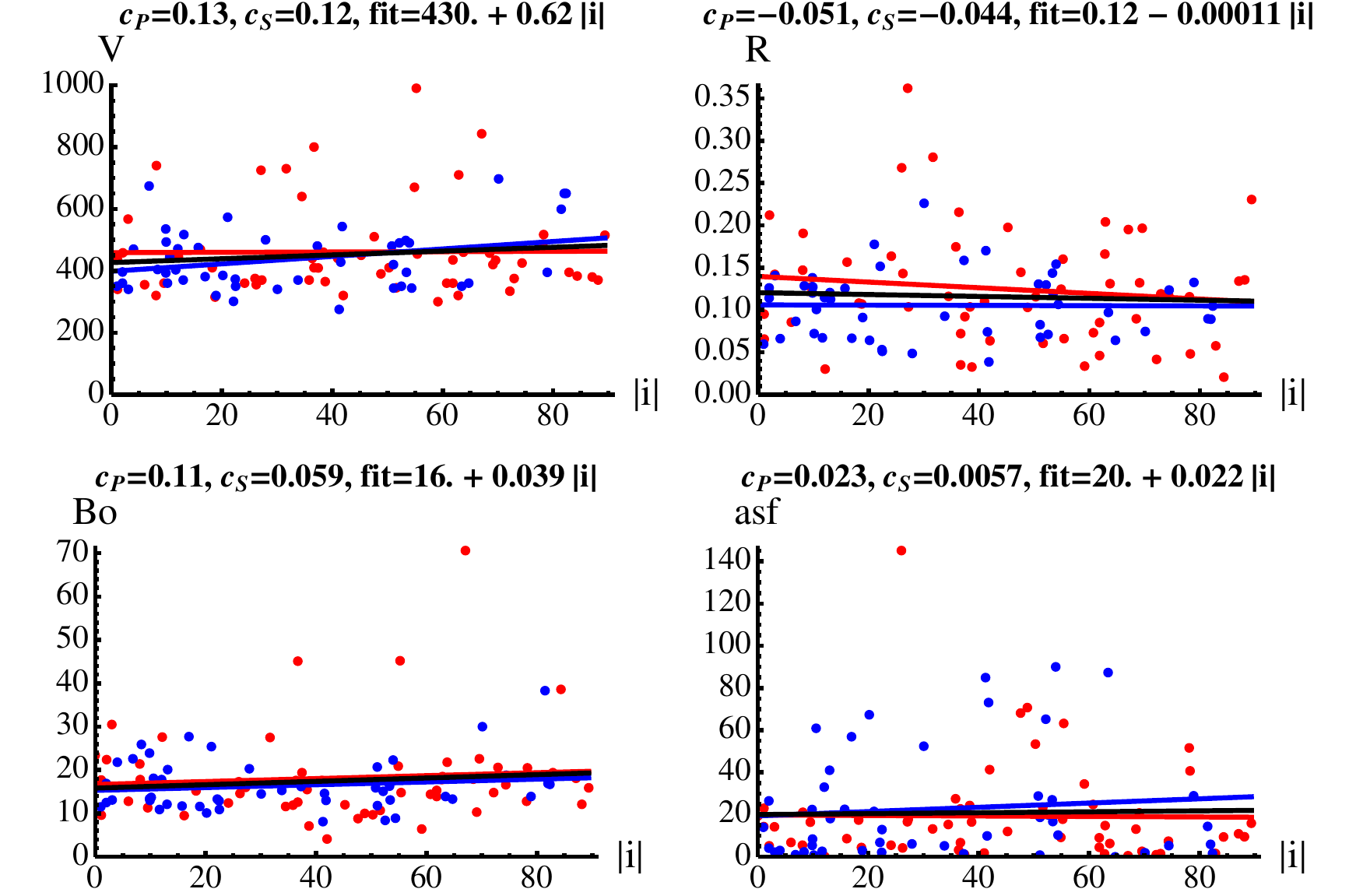}
 \caption{Properties of MCs observed at 1~AU versus the axis inclination on the ecliptic ($\iA$ in degree). The correlations are shown for: the mean MC velocity ($V$ in km/s), the MC radius ($R$ in AU), the axial magnetic field strength ($B_0$ in nT), and the asymmetry factor (asf in \%) for the full set of MCs.  The asymmetry factor \citep[asf, see][]{Lepping05} measures twice the time difference between the middle of the MC time interval and the closest approach (``center time''). It is expressed in \% of the MC event duration. It has been introduced to measure how far in time the peak in the modeled magnetic field is from the mid-point of observed MC. The drawing convention is the same as in \fig{corel_lambda}.}
\label{fig_corel_i}
\end{figure*}

\section{Observations } 
\label{sec_obs}

\subsection{Set of observed MCs} 
\label{sec_obs_Set}
  We first summarize the identification of MCs as defined by \citet{Lepping90}.  They first identified the time intervals having the four characteristics of MCs in the WIND data \citep[defined by][]{Burlaga81}.  Then, they determined the MC boundaries with the jumps in the plasma and magnetic field measurements and they fitted the magnetic field in the selected time intervals with a flux-rope model. This model assumes a linear force-free, or constant-$\alpha$, magnetic field \citep{Lundquist50}.   The least square fit to the \insitu\ data determines seven parameters of the model: (1) the longitude ($\pA$) and (2) the latitude ($\tA$) of the flux-rope axis (see \sect{obs_Axis}), (3) the distance of the spacecraft from the flux-rope axis at closest approach point ($Y_0$), (4) the magnetic field strength on the flux-rope axis ($B_0$),
(5) the twist ($\alpha$), (6) the sign of the magnetic helicity ($H = \pm 1$), and (7) 
the time at closest approach to the flux-rope axis ($t_0$).  The mean velocity ($V$) of the MC
is directly determined from the measured proton velocity.  From these parameters, other physical quantities of the flux-rope are computed, such as the flux-rope radius ($R$) and the impact parameter ($p=Y_0/R$). 

In the present study, we use an extended list of events (Table~2 at http://wind.nasa.gov/mfi/mag\_cloud\_S1.html) which is based on the results of \citet{Lepping10} and includes more recent MCs.
This list, at the date of February 13th 2013,
contains the parameters obtained for 121 MCs observed nearby Earth by WIND spacecraft from February 1995 to December 2009.
However, when removing the cases where the handedness could not be determined (flag f in the list) or the fitting convergence could not be achieved (flag F), this list restricts to 111 MCs. {Within the remaining cases}, 4 MCs have an impact parameter $p>1$ (so a fitted flux-rope extending beyond the first zero of the axial field in the Lundquist model).
Removing these suspicious cases, all of the worse class (quality 3, where the quality is defined in \citet{Lepping90} according to the $\chi^2$ value of the fit of a flux-rope model to data), 107 MCs remain, ranging from quality 1 to quality 3.

\subsection{Definition of the axis orientation} 
\label{sec_obs_Axis}

The WIND data are defined in the Geocentric Solar Ecliptic (GSE) 
system of reference (with unit vectors $\uvec{x}_{GSE}$, $\uvec{y}_{GSE}$, $\uvec{z}_{GSE}$), where 
$\uvec{x}_{GSE}$ points from the Earth toward the Sun, 
$\uvec{y}_{GSE}$ is in the ecliptic plane and in the direction opposite to the planetary motion, and $\uvec{z}_{GSE}$ points to the north pole.
The flux-rope axis orientation is classically defined in spherical coordinates by two angles: the longitude ($\pA$) and the latitude ($\tA$) as shown in \fig{schema}a.  The polar axis of the spherical coordinates ($\theta \approx 90$\degree) is singular as it corresponds to any values of $\pA$.  The above choice for a reference system sets this axis along $z_{GSE}$ which is both a possible and an un-particular axis direction. Therefore, the coordinates ($\pA,\tA$) are not appropriate to study the correlation of MC  properties with $\pA$ (we would need to limit the study to low $|\tA|$ values to have meaningful $\pA$ values).  

The Earth-Sun direction is a particular one considering the encounter of MCs coming from the Sun by a spacecraft.  We then set a new spherical coordinate system with its polar axis along $x_{GSE}$ (\fig{schema}a). {Since this direction corresponds in theory to the spacecraft crossing the flux rope parallel to the legs, and since in practice it is not possible to detect flux rope legs (\eg, the magnetic field rotation is very difficult to detect in the partial and longterm crossing of a leg), this direction does not appear in the MC data set studied here.} Then, we define the inclination on the ecliptic ($\iA$) and the location ($\lA$) angles (\fig{schema}).  {The names for these angles are derived from a MC with an axis located in a plane and with the distance to the Sun increasing along the flux-rope from any of its legs to its apex (as shown in \fig{schema}c)}. The angle $\iA$ is the inclination of this plane (in light blue) on the ecliptic (in light grey) as shown in 
\fig{schema}a,b.  The angle $\lA$ is evolving monotonously along the flux-rope, implicitly marking the location where the spacecraft intercepts the flux-rope (\fig{schema}c). {It defines} the position of the spacecraft crossing explicitly if the axis shape is known.

As for the latitude angle $\tA$, the inclination angle $\iA$ is defined in the interval $[-90 \degree,90 \degree]$ with $\iA=0$ when the MC axis is in the ecliptic plane (corresponding to $\tA=0$).  $\lA$ is measured from the plane ($\uvec{y}_{GSE},\uvec{z}_{GSE}$) towards the MC axis (\fig{schema}a).  By contrast, the cone angle $\bA$ was defined from $\uvec{x}_{GSE}$ towards  
the MC axis \citep[\eg][]{Lepping90}.  These angles are simply linked by $\lA = 90\degree -\bA$.  At the MC apex, $\bA \approx 90\degree$, while the MC legs have $\bA \approx 0\degree$ or $180\degree$.  It implies that $\bA$ is not a convenient angle to compare results on both sides of the apex since {the data cannot be reported on the same abscissa. However, choosing $\lA$ in $[-90\degree,90\degree]$ allows to do so, as it is shown by blue and red dots in \fig{corel_lambda}. }  
This is {why we} introduce the location angle {as a continuously changing quantity}: from $\lA \approx -90\degree$ in one leg, to $\lA \approx 0\degree$ at the apex, to $\lA \approx 90\degree$ in the other leg for a flux-rope axis having a curvature always directed inward (as in \fig{schema}c).  With this definition of $\lA$, the properties of both legs are simply compared by using |$\lA|$.  Next, if the flux-rope is not north-south oriented {(\eg\ with a plane close to the ecliptic plane),} then $\lA >0$ in the east leg and $\lA <0$ in the west leg (\fig{schema}c). In the case of a flux-rope more north-south oriented (inclined with the ecliptic plane), then $\lA >0$ and $\lA <0$ correspond respectively to the northern and southern legs for $\iA >0$ (and the reverse for $\iA <0$).   

The relations between ($\iA ,\lA$) and ($\pA ,\tA$) are simply:
\BA
\sin \lA &=& \cos \pA ~\cos \tA    \label{eq_lA} \,,\\
\tan \iA &=& \tan \tA ~/~ |\sin \pA|  \label{eq_iA} \,,
\EA 
where we include the absolute value of $\sin \pA$ since $\iA$ evolves similarly as $\tA$.

\subsection{Statistical properties of the axis orientation} 
\label{sec_obs_Statistical}
We analyze below the correlations {of the local axis orientation parameters with the other MC parameters deduced from the Lundquist model} for the set of 107 MCs. {The correlation analysis allows us to obtain proper sets of data to study the distribution of the local axis orientation parameters.}
 
{We present some of the correlation analysis results for the location angle $\lA$ in \fig{corel_lambda} and we find that} $\lA$ has only weak correlations with the other MC parameters (apart for $\pA$  and $\tA$ since the correlation is present from the definition, \eq{lA}).   A general result is also that there is no significant difference between both legs (\ie\ $\lA>0$ and $\lA<0$ as defined in \fig{corel_lambda}), so that in the following we only describe correlations with $|\lA|$.
For the full set of MCs, the strongest correlation is obtained with the flux-rope radius $R$ (\fig{corel_lambda}, top right).  Still, this correlation is quite weak regarding both the Pearson and Spearman rank correlation coefficients ($c_P =-0.29$, $c_S =-0.24$). Moreover, this correlation is even weaker 
($c_P = c_S =-0.12$) if we limit the analysis to a set with the best and good cases \citep[quality 1 and 2 as defined by][]{Lepping90}, \ie\ 74 MCs. The next significant correlation is with the impact parameter (not presented here, with $c_P =0.2$, $c_S =0.13$) but this small correlation almost vanishes for a set with only quality 1 and 2 MCs ($c_P \approx c_S \approx -0.03$). Then, the next largest correlation is between $|\iA|$ and $|\lA|$ (\fig{corel_lambda}, bottom right), and this weak correlation is kept with the quality~1 and 2 MCs set.  The other MC parameters show no significant correlation with $\lA$, \eg\ for $V$ and $B_0$ (\fig{corel_lambda}).  

The inclination angle, $\iA$, has even lower correlation with the other MC parameters compared to $\lA$ (\fig{corel_i}).  Very similar results are obtained with the sets $\iA>0$ and $\iA<0$ (by comparing red and blue points and lines in \fig{corel_i}). The best correlation is found with the MC velocity, $V$.  Still, it is a very weak correlation ($c_P =0.13$, $c_S =0.12$).  {To confirm the above results presenting very weak correlations between $\lA$, $\iA$ and the other MC parameters,} we further investigated the correlations obtained by ordering first the full MC set by growing value of one MC parameter \citep[such as $V$, $R$, $B_{0}$ and more generally all the parameters reported in the Table of][]{Lepping10}. Then, we computed the mean value of $\iA$ in subsets of MCs, scanning growing values of the selected parameter. This analysis (not presented here) confirmed that there is no significant dependence of $\iA$ with any of the other MC parameters (apart with $\tA$ and $\pA$ because of the definition, \eq{iA}). The present results for $\lA$ and $\iA$ imply that the MC properties are statistically independent of the axis orientation around the Sun-Earth line, as far as the limited number of MCs studied allows to conclude.  

\begin{figure}    
\centerline{\includegraphics[width=0.45\textwidth, clip=]{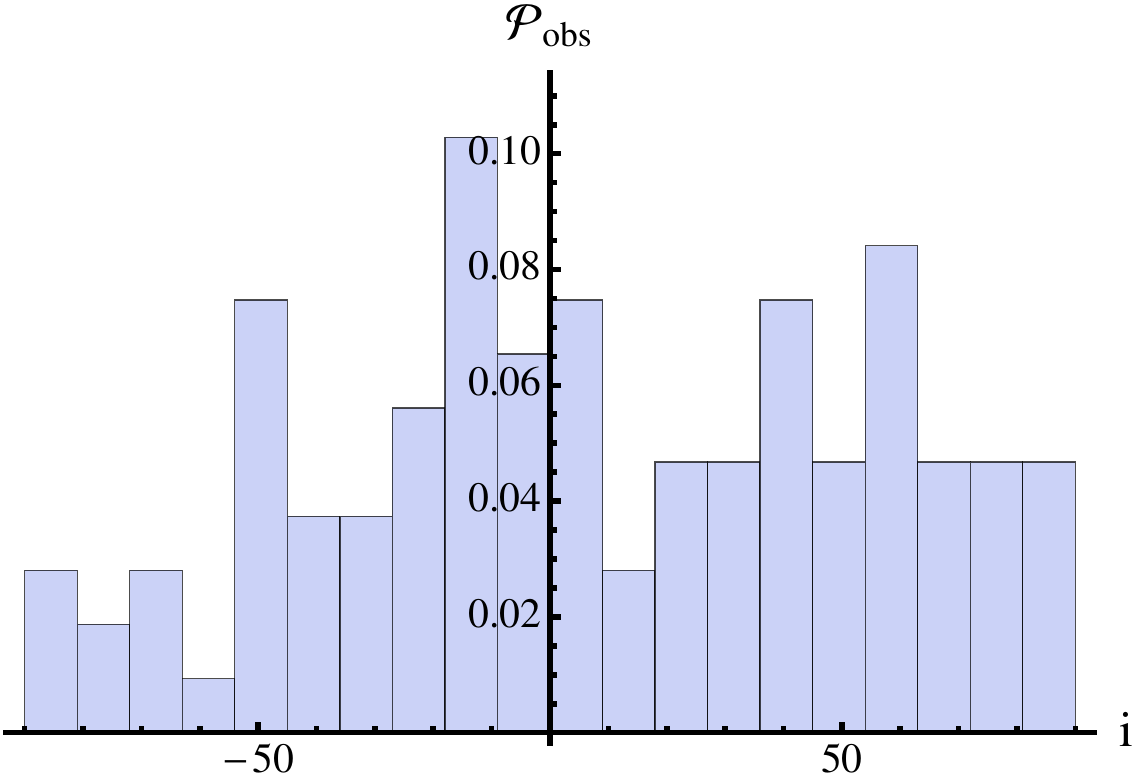}}
\caption{Probability distribution, $\pobs (\iA)$, of the inclination angle ($\iA$) as derived
from the magnetic data of 107 MCs observed at 1~AU and fitted by the Lundquist model \citep{Lepping90}. The data are grouped into a histogram having 20 bins of $\iA$. $\pobs$ is normalized so that the sum of the bins is unity.
}
 \label{fig_prob_i}
\end{figure}

\begin{figure}    
\centerline{\includegraphics[width=0.45\textwidth, clip=]{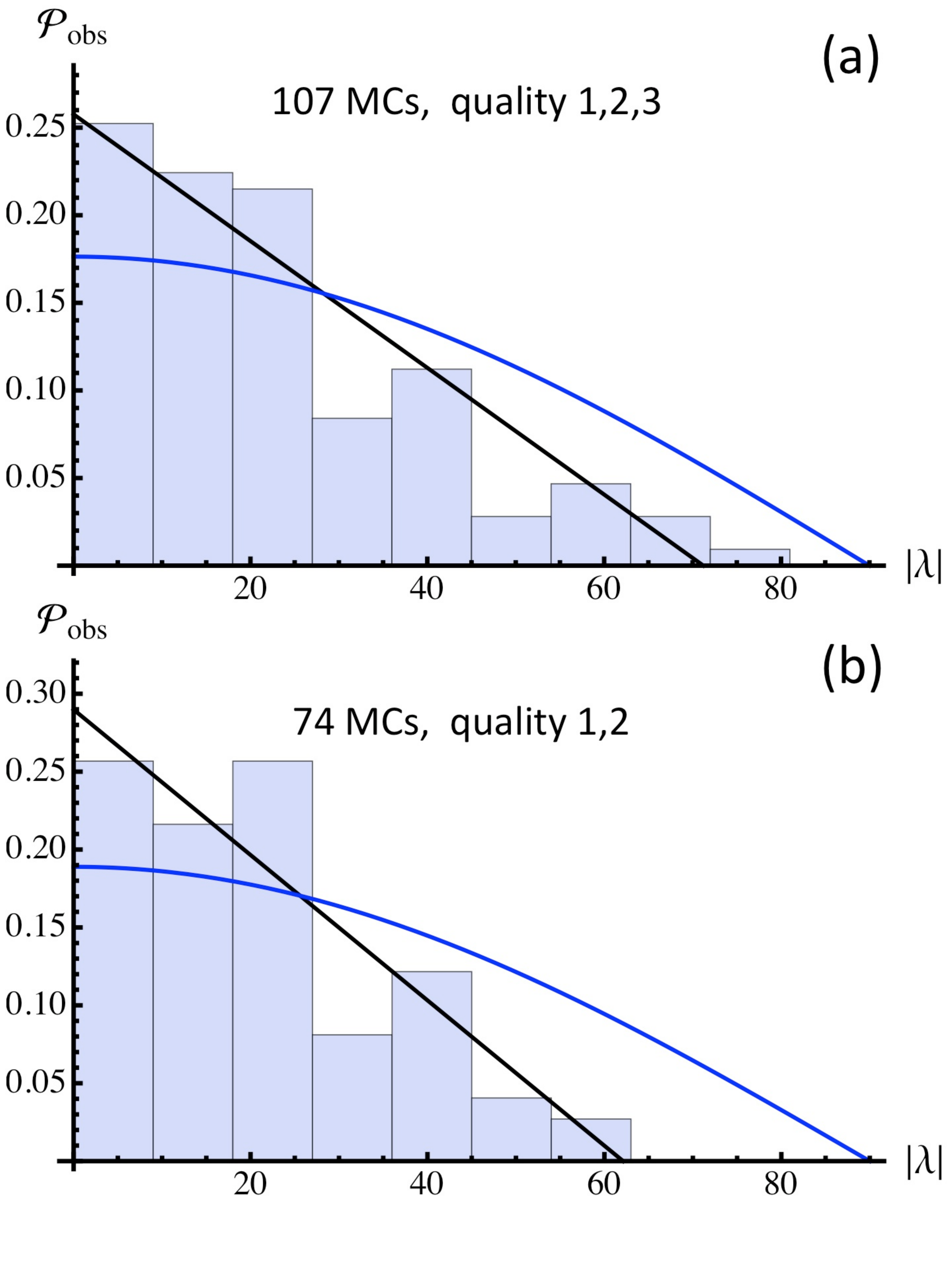}}
\vspace{-0.5cm}
\caption{Probability distribution, $\pobsl$, of the location angle ($\lA$) as derived
from the magnetic data of MCs observed at 1~AU and fitted by the Lundquist model \citep{Lepping90}. The data are grouped into a histogram having 10 bins of $|\lA|$ and $\pobsl$ is normalized so that the sum of the bins is unity.
A least square fit of the histogram with a straight line and a cosinus function are shown respectively in black and blue.
}
 \label{fig_prob_lambda}
\end{figure}

\begin{figure}    
\centerline{\includegraphics[width=0.45\textwidth, clip=]{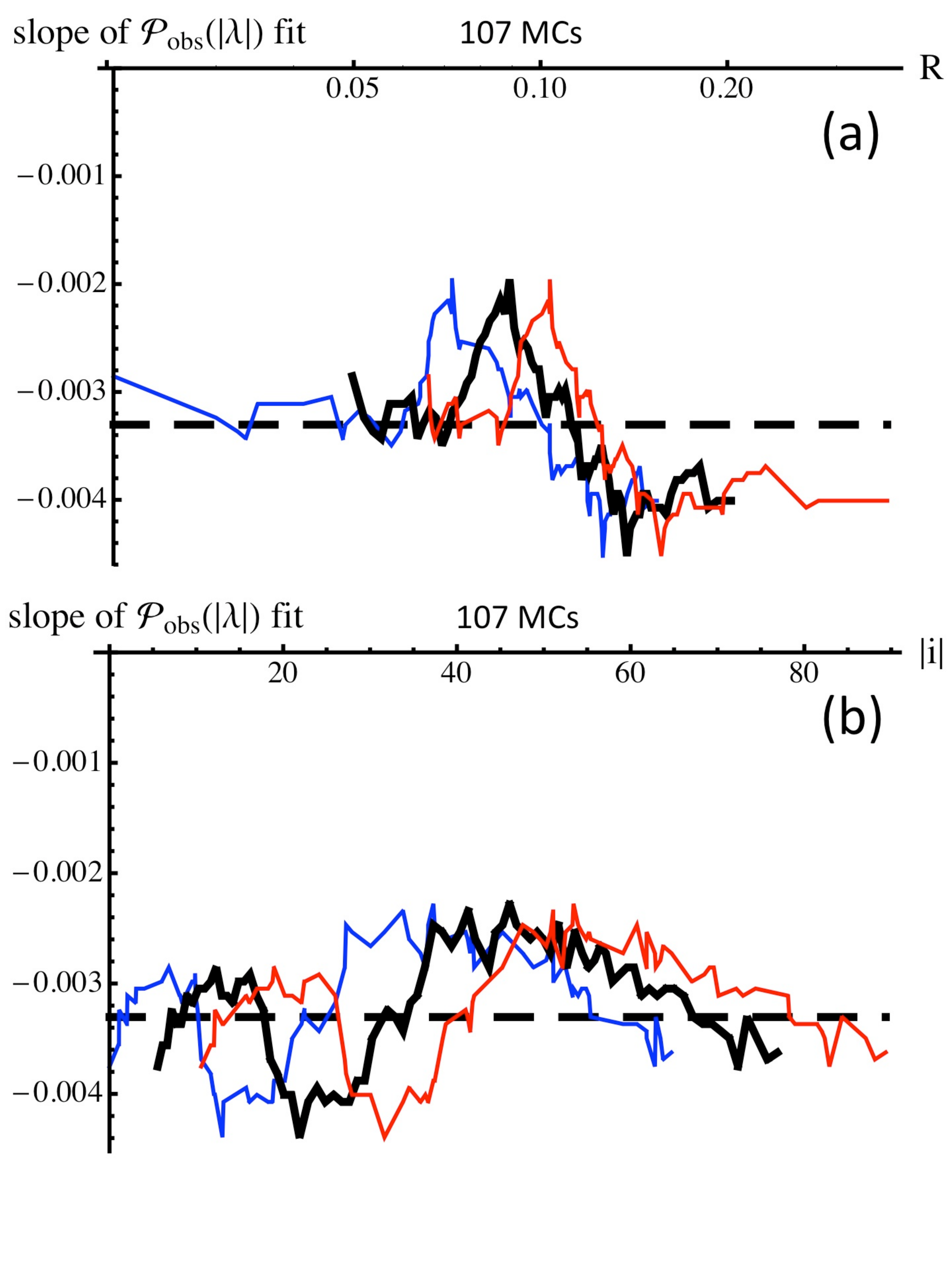}}
\vspace{-0.8cm}
\caption{Global property of the probability distribution $\pobsl$, parametrized by the slope of the linear fit (see the straight black line in \fig{prob_lambda}a).  This slope is shown here in function of two selected MC parameters. The MCs are first ordered by growing order of one parameter, then they are split in subsets of 20 MCs, shifting progressively the mean parameter to larger values.  
  The two selected parameters are the flux-rope radius ($R$ in AU, top panel) and the absolute value of the inclination angle ($|\iA|$, bottom panel).  
The three curves represent the slope of the fit, with the black line corresponding to the mean value of $R$ ($|\iA|$ for bottom panel) for each subset, and the blue (resp. red) line corresponding to the minimum (resp. maximum) of $R$ ($|\iA|$ for bottom panel) value for each subset. The horizontal dashed line is the slope for all MCs (black line slope in the top panel of \fig{prob_lambda}).  
}
 \label{fig_slopePlambda}
\end{figure}

\subsection{Distributions of the axis orientation} 
\label{sec_obs_Dist}

  The probability distribution of the axis inclination $\iA$, presented in \fig{prob_i}, is broad with flux-ropes {being detected} in all range of $\iA$, from $[-90\degree,90\degree]$.  The main maximum is for flux-ropes oriented close to the ecliptic plane, and there are secondary maximum for $|i|\approx 50\degree$.   There is also a marked difference with the sign of $\iA$: the cases with $\iA>0$ are nearly evenly distributed (within the statistical fluctuations) compare to those with $\iA<0$. Altogether, it implies that the flux-rope inclination on the ecliptic is broadly distributed without one strong privileged direction. 

  In contrast, the probability distribution of $\lA$ (\fig{prob_lambda}) is strongly non-uniform with a probability decreasing rapidly with growing $|\lA|$. \citet{Marubashi97} mentioned the 
possibility of finding the direction of the flux rope legs following the Archimedean spiral. From 
numerical simulations, \citet{Vandas02b} found a similar trend, finding evidences for an orientation of the legs similar to the solar wind Parker spiral. However, similar distributions are obtained for $\lA>0$ and $\lA<0$ (not shown) within the limit of statistical fluctuations, in particular for larger $|\lA|$ values (corresponding to MC legs where a low number of MCs are detected).  Restricting the MC set to the quality 1 and 2, so to 74 MCs, removes all the large $|\lA|$ values (\fig{prob_lambda}b).  It implies a distribution more peaked at low $|\lA|$ values.

  The cases with large $|\lA|$ correspond to the spacecraft crossing the region of a MC leg.  These cases typically lead to the largest uncertainty of the fitted flux-rope parameters \citep[\eg][]{Lepping10} {because of the difficulties in fitting a Lundquist model to the data.
They are observed in the regions of the MC legs and changing their locations in the distribution tail only weakly modifies the global distribution. As such, we consider the whole distribution of $\lA$ without truncating it.
As shown in \fig{prob_lambda},} a much stronger effect is present by selecting the MCs with the quality class.  Moreover, the low number of cases in the distribution tail implies that the $\lA$ distribution has large statistical fluctuations for large $|\lA|$ values. 
       
  We further study the probability distributions of $\lA$ by fitting them with a straight line (black line in \fig{prob_lambda}) in order to decrease the statistical fluctuations.  
The slope of the line, or simply slope of $\pobsl$, is directly linked to the mean of the distribution $<\!\!|\lA|\!\!>$ \citep[see Eq.~(15) and related text in][]{Demoulin13}.  
Since the statistical fluctuations of the mean value of $|\lA|$
are of the order of $<\!\!|\lA|\!\!>\!\!/\!\sqrt{N}$, where $N$ is the number of MCs in the distribution, this fitting procedure allows us to split the MC data set in subsets
while still keeping relatively low statistical fluctuations on the slope (see \fig{slopePlambda}).  It implies that we can test whether the probability distribution of $|\lA|$ is affected by some of the other MCs {parameters}. To do so, we first order the MCs by growing order of one selected {parameter}, and we then fit $\pobsl$ for each subset of $N$ MCs, progressively shifting to higher values of the selected MC parameter (by step of one MC). This allows the study of the slope evolution of the fit versus the selected parameter.

We find no significant dependence of the slope of $\pobsl$ with any of the other MC parameters except a weak one with the flux-rope radius {that we show in \fig{slopePlambda}a for subsets of $N=20$ MCs.}
Indeed, the smaller MCs ($R<0.1$~AU) have a slightly weaker slope, 
as they have a broader $\pobsl$ than the larger MCs.  A fluctuation of the slope of similar amplitude is also found when the MCs are ordered with $|i|$. Still, there is no significant slope difference between the MCs more parallel to the ecliptic (say $|i|<20\degree)$ from those more inclined on it.   We conclude that the probability distribution $\pobsl$ is almost independent of the orientation of the flux-rope, as well as of other MC parameters (not shown), except for the weak dependence on $R$.

The probability distribution of $\lA$ is closely linked with the mean shape of the axis (\fig{schema}c).  For example, with a circular shape and relatively close legs (separated by an angle of less than few $10\degree$), the expected $\lA$ distribution is $f_{\rm cos} = c \cos |\lA|$ (where c is a constant for normalizing the total probability to 1), as will be shown and discussed in detail in \sect{Simple_Particular}.  The fit of $f_{\rm cos}$ to the observed distribution, as shown by the blue curve in \fig{prob_lambda}, indicates that $f_{\rm cos}$ is a too broad distribution compared to the observed one.  This difference is even stronger in the case where only the MCs of quality 1 and 2 are considered.  This already implies that the axis shape is flatter than a circular one.

\begin{figure}[t!]    
\centering
\includegraphics[width=0.4\textwidth, clip=]{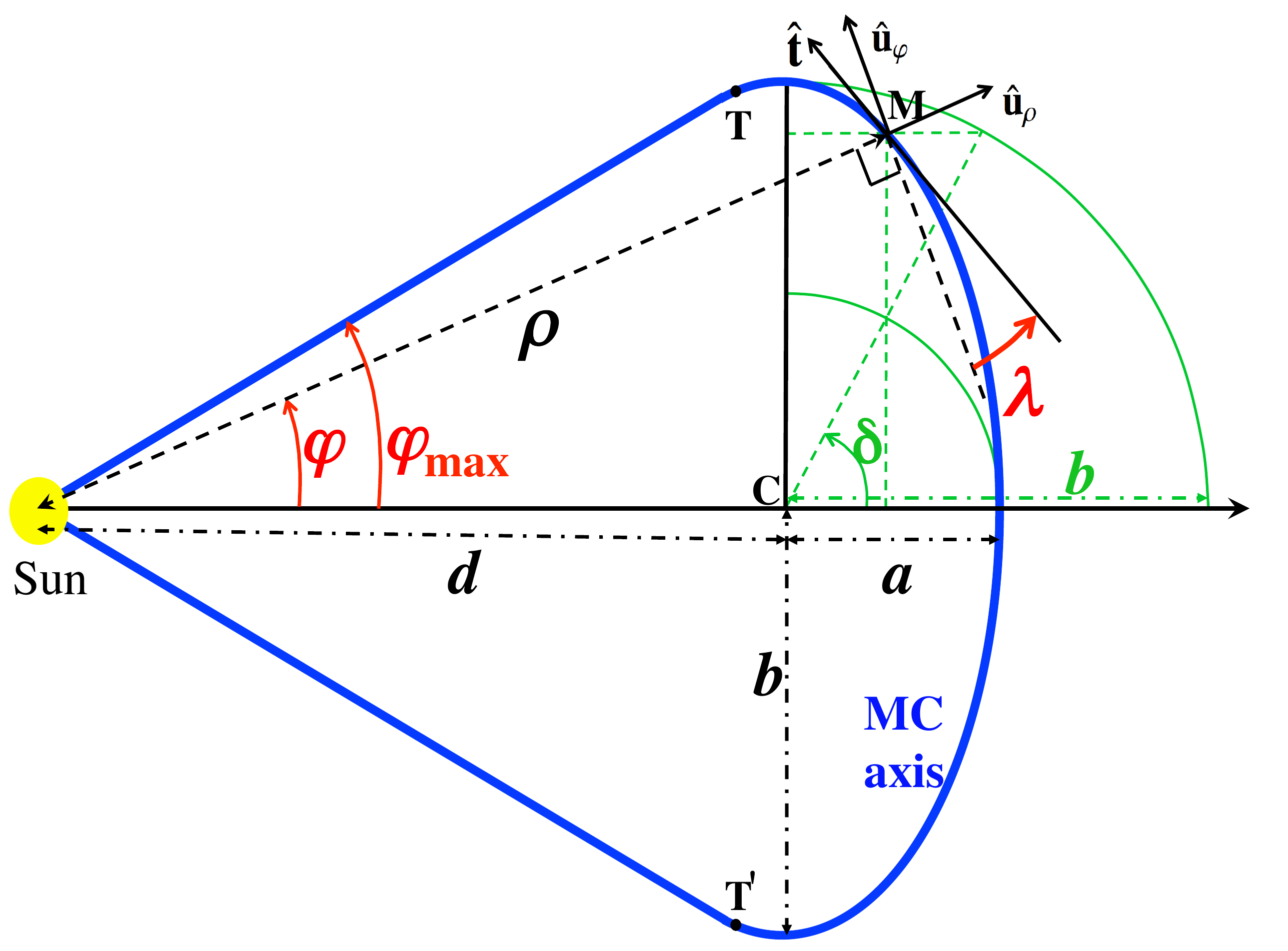}
\caption{Schema defining a model of the flux-rope axis with an elliptical shape.
The legs are represented by straight and radial segments tangent to the ellipse {and linking it to the Sun}.  $\varphi$ is the angle of the cylindrical coordinates ($\rho,\varphi$). The point C is the ellipse centre and M is the point of interest (where the spacecraft crosses the flux-rope).
}
 \label{fig_schema_E}
\end{figure}

\section{Simple models of {a global flux rope} axis} 
\label{sec_Simple}

Since the derivation of the mean shape of the axis from $\pobsl$ is not fully straightforward, we analyze in this section the reverse problem, \ie\ computing the location angle distribution from given models of the global axis shape. In particular, how sensitive is the $\lA$ probability distribution to the global axis shape and what are the main axis {geometry} parameters affecting the distribution?
     
\begin{figure*}
\centering
\sidecaption    
\includegraphics[width=12cm]{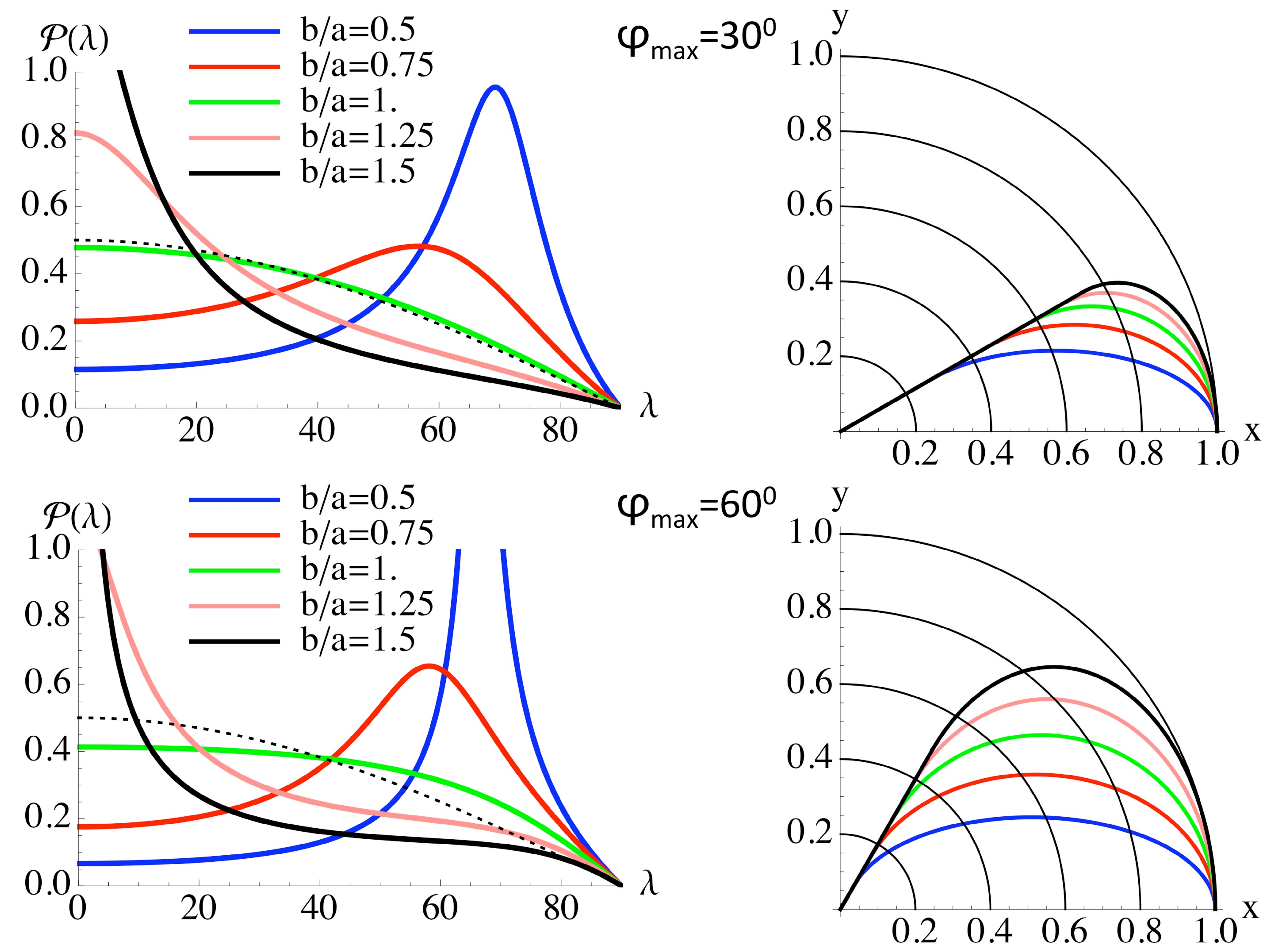}
 \caption{
Probability distribution of the location angle $\lA$ (left panels) and the corresponding axis shape (right panels) for the elliptical model of the flux-rope axis (defined in \fig{schema_E}).  
Since the model is symmetric, $\mathcal{P}(-\lambda) = \pl$ so that only the part $\lA > 0$ is shown. The distributions are normalized so that the integral of $\pl$ is unity. Five cases with different values of the aspect ratio ($b/a$) of the ellipse are shown in solid curves, for two maximum extension $\phimax$ of the angle $\varphi$ (defined in \fig{schema_E}).
The dotted curve represents the distribution for a circular front and small $\phimax$ values (cosinus function, see \sect{Simple_Particular}).
}
\label{fig_E_bSa}
\end{figure*}

\subsection{Axis model with an elliptical shape} 
\label{sec_Simple_Elliptical}

  We select a flux rope model which has enough free parameters to describe a large variety of axis shapes,
but which also has a minimum of complexity needed.  The flux-rope axis is supposed to be planar and it is described by a portion of an ellipse, up to the points T and T', where the tangent to the ellipse is {a radial segment attached to the Sun} (\fig{schema_E}).  These straight segments simply describe the flux-rope legs and link the flux-rope to the Sun. As such, the ellipse is not directly attached to the Sun as \eg\ in \citealt{Krall07}.  
{Note that the in situ measurements in a region within the flux-rope legs with $\lA \approx 90$\degree} do not show an important rotation of the magnetic field while the spacecraft crosses the MC. As such, these events are typically not reported as MCs \citep[][and references therein]{Owens12}. Similarly here, we do not consider the straight parts of the axis model in the computed distributions of location angles $\lA$.  

The ellipse centre, C, is at a distance $d$ from the Sun and its axis are along the radial and ortho-radial directions with half size $a$ and $b$, respectively (\fig{schema_E}).  
A point M on the elliptical part of the axis is at a distance $\rho$ from the Sun:  
  \BE \label{eq_E_rho}
  \rho = \sqrt{(d+a \cos \delta)^2 +  (b \sin \delta)^2} \, ,
  \EE
where $\delta$ is the angle defining the position of M from the ellipse centre.  The other cylindrical 
coordinate of M, $\varphi$, is given by
  \BE \label{eq_E_varphi}
  \tan \varphi = b  \sin \delta /(d+a \cos \delta) \, .
  \EE

The angle between the tangent to the ellipse at M and the local ortho-radial direction from the Sun is the location angle $\lA$ (\fig{schema_E}).
It is related to the other angles and ellipse parameters by  
  \BE \label{eq_E_tanLambda}
  \tan \lA = \frac{a \sin \delta \cos \varphi - b \cos \delta \sin \varphi}
                  {a \sin \delta \sin \varphi + b \cos \delta \cos \varphi} \, .
  \EE
The above equation is simplified by the introduction of the angle $u$ defined as
  \BE \label{eq_E_u}
  \tan u = \frac{a}{b} \tan \delta \, .
  \EE
Then, \eq{E_tanLambda} simplifies to
  \BE \label{eq_E_Lambda}
  \lA = u-\varphi \,,
  \EE
with $\lA$ within the interval $[-90\degree,90\degree]$.
This equation expresses implicitly the angle $\lA$ in function of $\varphi$ and the parameters
$(a,b,d)$ after eliminating $u$ with \eq{E_u} and $\delta$ with \eq{E_varphi} (\ie\ expressing $\cos \delta$ in function of $\tan \varphi$). 

In summary, this elliptical model of the axis depends on three parameters: $\{a,b,d\}$.
Equivalently, it is also defined by these three other parameters: $\{a+d,b/a,\phimax \}$
representing respectively the apex distance from the Sun, the aspect ratio and the maximum angular extension defined by:
  \BE \label{eq_varphi_max}
  \tan \phimax = b / \sqrt{d^2-a^2}  \, .
  \EE

\begin{figure*}
\centering
\includegraphics[width=\textwidth]{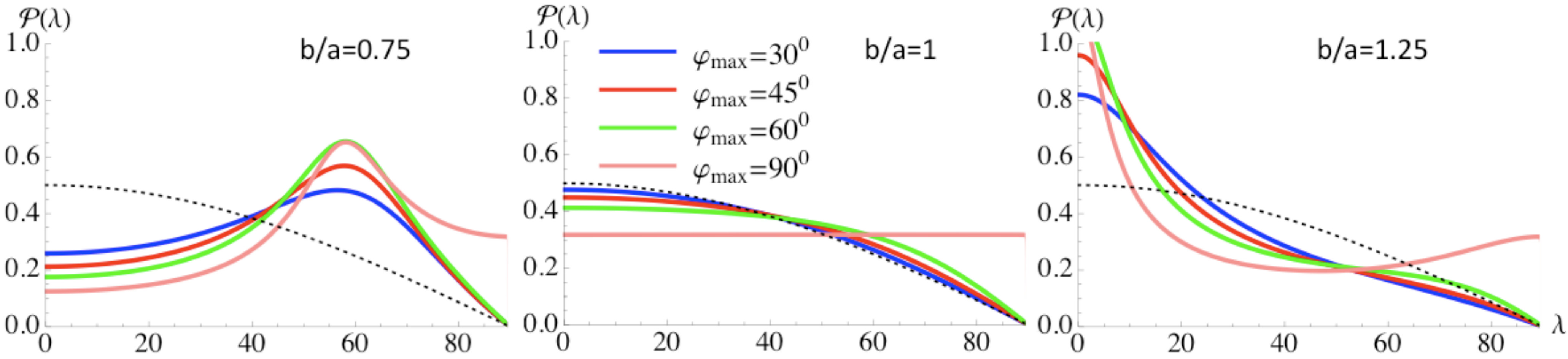}
 \caption{
Comparison of the probability distribution of the location angle $\lA$ for the elliptical model of the flux-rope axis with a given aspect ratio $b/a$ in each panel (see \fig{schema_E}). The distributions are normalized as in \fig{E_bSa} and the dotted curve is the distribution for a small circular front. The maximum extension angle $\phimax$ has only a weak effect on the distribution shape compare to the large effect of the aspect ratio $b/a$.
}
\label{fig_E_phi}
\end{figure*}

\subsection{Probability distribution of the location angle $\lA$} 
\label{sec_Simple_Location}

The \insitu\ observation of a MC made from a single spacecraft only provides a local estimation of the flux-rope axis orientation (by fitting a flux-rope model to the magnetic field data).  So for a given MC,
only one value of $\lA$ is available, say at point M along the flux rope axis (\fig{schema_E}). Let us first consider a series of flux-ropes contained in the ecliptic plane (\ie\ $\iA \approx 0$).
On the time scale of a solar cycle, the Sun is launching MCs from any longitude. Moreover, since the Sun is rotating, any privileged active longitude is covered over a time scale of $\sim$ 11 years.  It implies that MCs are expected to be observed with an equiprobability of $\varphi$, except for an expected lower rate of detection in the legs due to an observational bias 
\citep[as the flux-rope is only partially crossed so it is not always detected, \eg][and references therein]{Owens12}.
     
 We have shown in \fig{prob_i} that a large fraction of flux-ropes is significantly inclined with respect to the ecliptic plane. However, since we find no significant correlation between any of the estimated flux-rope characteristics and the angle $\iA$ (\sect{obs_Statistical}), it is reasonable to expect an equiprobable distribution of $\varphi$ for any $\iA$ angle. 
Moreover, the probability distribution of $|\lA|$ remains similarly peaked towards low $|\lA|$, with a mean slope almost independent of $|\iA|$ (\fig{slopePlambda}b).  We deduce that the range of solar-latitude launch is broad enough to allow a similar scan of flux-rope axis with significant $|\iA|$ values as the ones with low $|\iA|$ values.   

 Then, we suppose that the probability of $\varphi$, $\pvphi$, is uniform within the set of detected MCs and for the above axis model, at least away from the legs. With $\varphi$ in the interval $[-\phimax , \phimax ]$, the distribution $\pvphi$ is simply a constant defined by the normalization of the total probability to unity:
  \BE \label{eq_pvphi}
  \pvphi = 1 / (2 ~\phimax ).
  \EE
 
For flux-ropes having an axis curved inward, as in \fig{schema_E}, there is a monotonous relationship between $\lA$ and $\varphi$.  
Considering that the intervals $[\varphi,\varphi +\rmd \varphi]$ and $[\lA,\lA +\rmd \lA]$ contain the same number of cases, we link the two probabilities $\pl$ and $\pvphi$ by
  \BE \label{eq_pl}
  \pl = \pvphi ~|\rmd \varphi/\rmd \lA | .
  \EE
With $\pvphi$ known and $\rmd \varphi/\rmd \lA$ computed from the equations of the above axis model, \eq{pl} provides the probability distribution of $\lA$
which can be compared with the observed ones (\fig{prob_lambda}).

The computation of $\rmd \varphi/\rmd \lA$ is realized by differentiating Eqs.~(\ref{eq_E_varphi}), (\ref{eq_E_u}) and (\ref{eq_E_Lambda}).  Regrouping these equations provides:
  \BA 
  \frac{\rmd \lA}{\rmd \varphi} = -1 
  &+& \frac{1+\tan^2 \delta}{1+ (a/b)^2\tan^2 \delta} ~~
    \frac{a}{b \cos \varphi}  \nonumber \\
  && \times \frac{d+a \cos \delta}{a \sin \delta \sin \varphi + b \cos \delta \cos \varphi} \, .
     \label{eq_E_dbetaSdLambda} 
  \EA
Then, $\pl$ is computed from Equations \eq{pvphi}, \eq{pl}, and \eq{E_dbetaSdLambda}.

\subsection{Particular probability distributions of $\lA$} 
\label{sec_Simple_Particular}

The probability $\pl$ has a simple expression at the apex (where $\lA =0$, $\rho=\rho_{\rm max}=a+d$, and $\varphi =0$)
  \BE \label{eq_pl_apex}
  \prob (\lA =0) = \pvphi ~ \frac{b^2}{|ad+a^2-b^2|} \, .
  \EE
Introducing the radius of curvature of the ellipse, $R_{\rm c}=b^2/a$, at $\varphi =0$,
\eq{pl_apex} is rewritten as
  \BE \label{eq_pl_apexRc}
  \prob (\lA =0) = \pvphi ~ \frac{1}{|\rho_{\rm max}/R_{\rm c}-1|} \, .
  \EE
It shows that $\pl$ becomes singular (infinite) at the apex when $R_{\rm c}=\rho_{\rm max}$, \ie\ when the ellipse is tangent to the circle $\rho=\rho_{\rm max}$, so that the front is locally the flattest possible (in cylindrical coordinates).
The black solid curves in \fig{E_bSa} illustrate cases with $R_{\rm c}$ very close to $\rho_{\rm max}$.{The left panels show the probability distributions that become infinite for $\lA \rightarrow 0$\degree\ (apex) for $\phimax =30 \degree$ and $60 \degree$, while the right panels show the half ellipse shape following the circle $\rho=\rho_{\rm max}$ near the apex (especially the case $\phimax =60 \degree$). Other cases with $R_{\rm c}$ close to $\rho_{\rm max}$ (\ie\ pink solid curves in \fig{E_bSa})} show that the corresponding probability $\pl$ is much more peaked at $\lA=0$ than the probability deduced from observations (\fig{prob_lambda}), even if we include the same binning (not shown).

The expression of the probability $\pl$ can also be simplified
in the limit of a circular front, \ie\ when $b=a$, as 
  \BE \label{eq_pl_circle}
  \pl = \pvphi ~\frac{a}{d}\frac{\cos \lambda}
                 {\sqrt{1 -(a/d)^2 \sin^2 \lambda}} .
  \EE
This result shows that for a narrow angular extension, \ie\ $a/d \ll1$ or equivalently $\phimax \ll 90\degree$, $\pl$ simply has a $\cos \lambda$ dependence.  Such $\pl$ is more extended in $\lA$ compared with observations (see the blue curves in \fig{prob_lambda}).  On the contrary, for a broad angular extension ($a=d$ or $\phimax =90\degree$, using \eq{varphi_max}), $\pl$ is uniform in $\lambda$ and $\pl = \pvphi$.  This case corresponds to an axis located on a circle attached to the Sun (so without the straight segments departing from T and T' in \fig{schema_E}).  Such a distribution is also incompatible with the distribution deduced from the observations (\fig{prob_lambda}).

\subsection{Expected probability distributions of $\lA$} 
\label{sec_Simple_Expected}

The elliptical model of the axis shown in \fig{schema_E} has three free parameters: $a,b$ and $d$.
We fix the global scale of the model by normalizing the sizes by $\rho_{\rm max}$, \ie\ fixing $a+d=1$. We explore below the effect of the aspect ratio $b/a$ and $\phimax$ on $\pl$, setting $\pvphi$ to a uniform distribution. 
The aim is to compare the variety of the computed distributions $\pl$ with the observed ones (\fig{prob_lambda}). 

For a given $\phimax$ value, the aspect ratio $b/a$ has an important effect on $\pl$ as shown in \fig{E_bSa}.
For $b/a=1$ (green curve), corresponding to a circular shape of the flux rope axis, $\pl$ is close to a $\cos \lA$ function except when $\phimax$ is getting close to $90\degree$, in agreement with \eq{pl_circle}.  As $b/a$ is slightly lower than $1$, $\pl$ deviates significantly from the $\cos \lA$ function with a peak appearing in the leg part (more precisely around $\lA \approx 60$-$70\degree$ for the blue and red curves) and growing rapidly as $b/a$ decreases (\fig{E_bSa}). In parallel, an important decrease of $\pl$ is present for $\lA \leq 50\degree$, therefore for a region near the apex region.  On the contrary, increasing $b/a$ above $1$ increases sharply the probability in the apex region at the expense of the leg region (pink and black curves, \fig{E_bSa}).

The above effect of $b/a$ is enhanced for larger $\phimax$ (lower panels of \fig{E_bSa}).   However, the effect of $\phimax$ is much lower than the effect of $b/a$ as shown in \fig{E_phi}.  Indeed, $b/a$ is the main parameter which defines the shape of $\pl$ with the presence of a peak in the leg region for $b/a<1$ and at the apex for $b/a~>~1$.  $\phimax$ only weakly modulates this main tendency and it has a significant effect on $\pl$ only for $\phimax$ close to $90\degree$ (so for an axis shape close to an ellipse directly attached to the Sun). 

{The comparison of the results for the distribution shown in \figs{E_bSa}{E_phi} with those obtained for the observations and shown in \fig{prob_lambda} reveals that the observed $\lA$ probability distribution sets stringent conditions for a global flux rope axis model}.  The axis shape needs to be flatter than a circular shape, but it cannot be too flat.
In particular, an aspect ratio $b/a$ of only $1.25$ (\fig{E_phi}) implies an already too peaked $\pl$ distribution around the apex compared to \fig{prob_lambda}. 

We conclude that, within the hypothesis of a uniform $\pvphi$ distribution and comparable axis shape for MCs, the observed distribution $\pl$ sets a stringent constrains on the mean axis shape.

\begin{figure*}
\centering
\includegraphics[width=\textwidth]{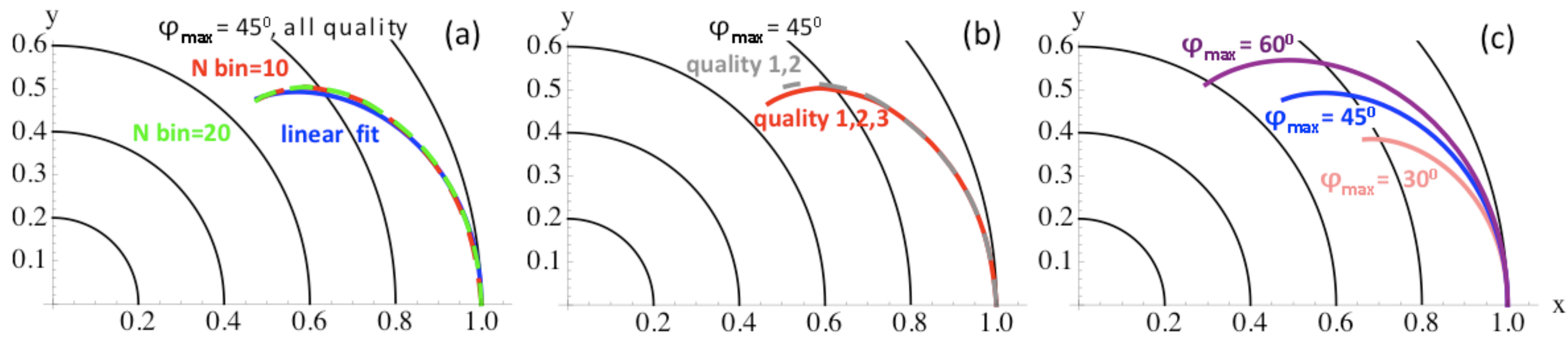}
 \caption{Mean flux-rope axis deduced from the probability distribution $\pobsl$ shown in \fig{prob_lambda}.
{\bf a)} Comparison of the axis deduced from a linear fit, and from a spline interpolation of $\pobsl$ with 10 and 20 bins for 107 MCs. 
{\bf b)} Comparison of the axis deduced directly from $\pobsl$ with 10 bins for all and quality 1,2 MCs, so from the two distributions shown in \fig{prob_lambda}.
{\bf c)} Effect of changing the free parameter $\phimax$. 
}
\label{fig_axis}
\end{figure*}

\section{Deduction of the axis shape from the data} 
\label{sec_Deduction}

The forward modeling presented in \sect{Simple} has emphasized the relationship between the shape of the flux-rope axis and the expected probability distribution $\pl$.  It specifies qualitatively which kind of axis shapes are closer to observations.   However there are still significant differences between the modeled $\pl$ distributions (\figs{E_bSa}{E_phi}) and the observed ones (\fig{prob_lambda}).
Rather than finding the optimum values of $b/a$ and $\phimax$ of the elliptic model that best fit the observed distributions, we derive below a procedure to obtain the axis shape from the observed $\pobsl$ distributions. 

\subsection{Method} 
\label{sec_Deduction_Method}

Similarly to \sect{Simple}, we suppose that the flux-rope axis of any analyzed MC is located in a plane
inclined by an angle $\iA$ on the ecliptic plane (\fig{schema}).  As such, we do not consider non-planar MC axis, as suggested by \citet{Farrugia11} for the observations of one MC
by three spacecraft. This would require a statistical analysis of the impact of deformed
axis on the probability distribution of $\lambda$, which is out of the scope of this paper. However, as was shown in \sect{obs_Dist}, since we find that the observed distribution $\pobsl$ is nearly independent of $\iA$ (\fig{slopePlambda}b), we can then suppose that the axis shape is independent of $\iA$. In the following, we only provide results derived from $\pobsl$ as shown in \fig{prob_lambda}. Next, we describe the flux-rope axis with the cylindrical coordinates ($\rho,\varphi$), as defined in \fig{schema_E}. 

From the Sun (the origin of coordinates), the {distance to the} M point on the MC axis can be expressed with the radius vector:
  \BE \label{eq_D_OM}
  \vec{SM} = \rho (\varphi) ~\ur  \, .
  \EE
We also suppose that $\rho$ is a decreasing function of $|\varphi|$ from the axis apex to any of the legs.   More precisely, we suppose that $\lambda$ is a monotonous function of $\varphi$ with $\lambda$ growing from $-90\degree$ to $90\degree$ as $\varphi$ evolves from $-\phimax$ to $\phimax$ (\fig{schema}c).
   Since we find no indication of an asymmetry between the legs in the MC data (\sect{obs_Dist}), we suppose $\rho(-\varphi)=\rho(\varphi)$, and we present the results only for $\pobsl$.
   Apart from these general constraints, the flux-rope shape is not prescribed, contrary to \sect{Simple}, and we deduce it from the observed distribution $\pobsl$ shown in \fig{prob_lambda}.      

The conservation of the number of cases implies that the variation of $\varphi$ is linked to those of $\lA$ as in \eq{pl} by:
  \BE \label{eq_D_dvarphi}
  \rmd \varphi = \frac{\pobsl}{\pvphi} ~\rmd \lA  \, ,
  \EE
and we suppose that $\pvphi$ is uniformly distributed in the interval $[0,\phimax ]$, so that $\pvphi = 1/\phimax $.  The integration of \eq{D_dvarphi} provides $\varphi$ as a function of $\lA$ as 
  \BE \label{eq_D_varphi}
  \varphi (\lA) = \phimax \int_{0}^{\lA} \pobs (|\lA'|) \rmd \lA'  \, ,
  \EE
with $\lA \geq 0$,  $\varphi \geq 0$.

Next, we relate $\rho$ to $\lA$.  Making the derivation of \eq{D_OM} with respect to $\varphi$, the unit tangent vector at point M is  
\BE \label{eq_D_t}
  \uvec{t} = \left(\frac{\rmd \ln \rho}{\rmd \varphi} \ur +  \, \up \right) 
              \scalebox{2}{/} \sqrt{1+\left(\frac{\rmd \ln \rho}{\rmd \varphi} \right)^2 } \, .
  \EE

The location angle $\lA$ is related to $\rho (\varphi )$ as 
  \BE \label{eq_D_tanLambda}
  \tan \lA = \frac{-\uvec{t}\cdot\ur}{\uvec{t}\cdot\up} 
           = -\frac{\rmd \ln \rho}{\rmd \varphi}   \, .
  \EE

Using \eq{D_dvarphi} together with $\pvphi = 1/\phimax $,  the integration of \eq{D_tanLambda} implies
  \BE \label{eq_D_rho}
  \ln \rho (\lA) = -\phimax  
            \int_{0}^{\lA} \tan (\lA') ~\pobs (|\lA'|) ~\rmd \lA'
           +\ln \rho_{\rm max}  \, .
  \EE

In summary, \eqs{D_varphi}{D_rho} provide $\varphi$ and $\rho$ as functions of $\lA$, so that the axis shape can be derived as a parametric curve in cylindrical coordinates.  It depends on two integration constants $\phimax $ and $\rho_{\rm max}$.  The second one is only a global scaling of the axis shape, which can be set to 1~AU for the application to WIND data. However, $\phimax$ is an intrinsic freedom parameter of the method and it is not defined by the \insitu\ observations.  Note that we found in \sect{Simple_Expected} that $\phimax$ has a small effect on the derived $\pl$ for the forward modeling of the axis with an elliptical shape (\fig{E_phi}). This is consistent with the present results since we have found here that the observed distribution $\pobsl$ is compatible with a large range of $\phimax$ values. 

Finally, we derive \eqs{D_varphi}{D_rho}. Since they involve integrals of the observed distribution $\pobsl$, the derived axis shape is expected to be weakly affected by the details of $\pobsl$.  Also, because the integration advances from the apex toward the legs, the growing uncertainties on $\pobsl$ with $|\lA|$ are kept for growing values of $|\lA|$, \ie\ the axis shape is expected to be best determined around the apex and with a growing uncertainty when going towards the legs.

\subsection{Results} 
\label{sec_Deduction_Results}

The values for the distribution $\pobsl$ {appearing in \eqs{D_varphi}{D_rho} can be computed} in different ways to deduce the mean axis shape. First, $\pobsl$ can be interpolated by Hermite or spline polynom functions before performing the integrations.  We set $\pobs (90\degree )=0$, and use the symmetry $\pobs (-\lA)=\pobs (\lA)$ in order to have an interpolation (and not an extrapolation) for all the $\lA$ ranges.  First, we find negligible differences between these two types of interpolation and between different interpolation orders (1 to 3).
Second, we use different numbers of bins to build the $\pobsl$ histogram.  The number of bins also has a negligible effect on the axis shape, as shown for 10 and 20 bins in \fig{axis}a. Finally, $\pobsl$ can be fitted by an analytical distribution.  The result with a linear function, as shown in \fig{prob_lambda} (black line), provides a very similar axis shape, as is shown in \fig{axis}a.  Such result also holds with other fitting functions such as a Gaussian distribution function.  All these tests confirm that the results derived from \eqs{D_varphi}{D_rho} are robust, \ie\ weakly affected by the local variations of $\pobsl$.      

With the definition of the quality class according to \citet[][]{Lepping90}, we keep only the best and good cases, so quality 1 and 2. Then, $\pobsl$ is more peaked near low $|\lA|$ values (\fig{prob_lambda}b).  It implies a slight change of the derived axis shape only far from the apex (\fig{axis}b), while the corresponding $\pobsl$ have more differences (\fig{prob_lambda}).   This is again an effect on the integration present in both \eqs{D_varphi}{D_rho}.   

Finally, the main uncertainty on the axis shape is due to the free parameter $\phimax$.  Indeed its effect is significant, especially away from the apex (\fig{axis}c).  However, for the expected range of  $\phimax$, as shown, the results imply that the axis is significantly bent, \ie\ more bent that the curvature of a circle of radius $\rho(\varphi) = \rho(\varphi=0)$. The axis shape is also slightly elongated in the ortho-radial direction (along $\up$), in agreement with an aspect ratio $b/a$ slightly above 1 when the axis is modeled with an elliptical shape (\sect{Simple}).

\begin{figure*}
\centering
\sidecaption    
\includegraphics[width=5.95cm]{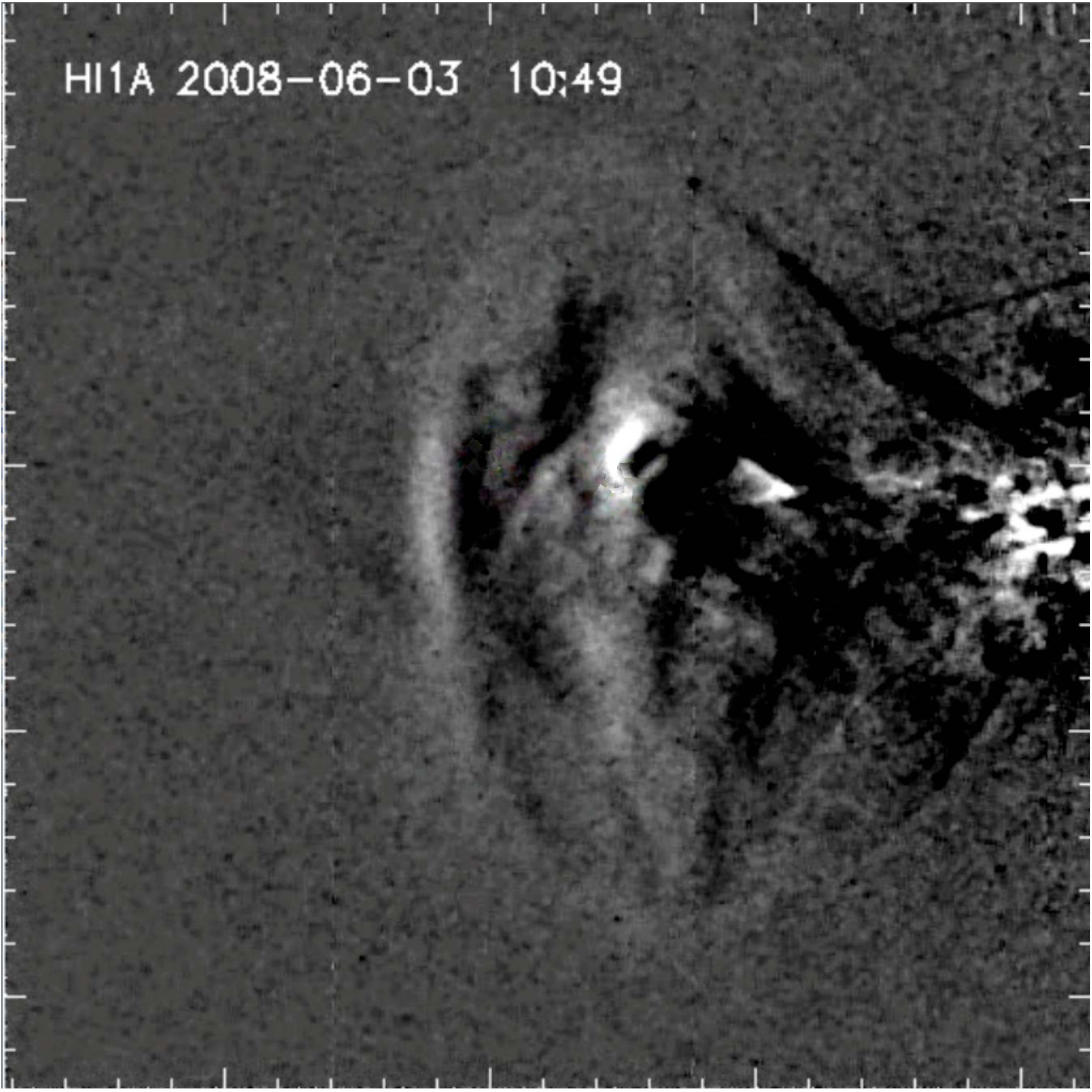}
\includegraphics[width=5.95cm]{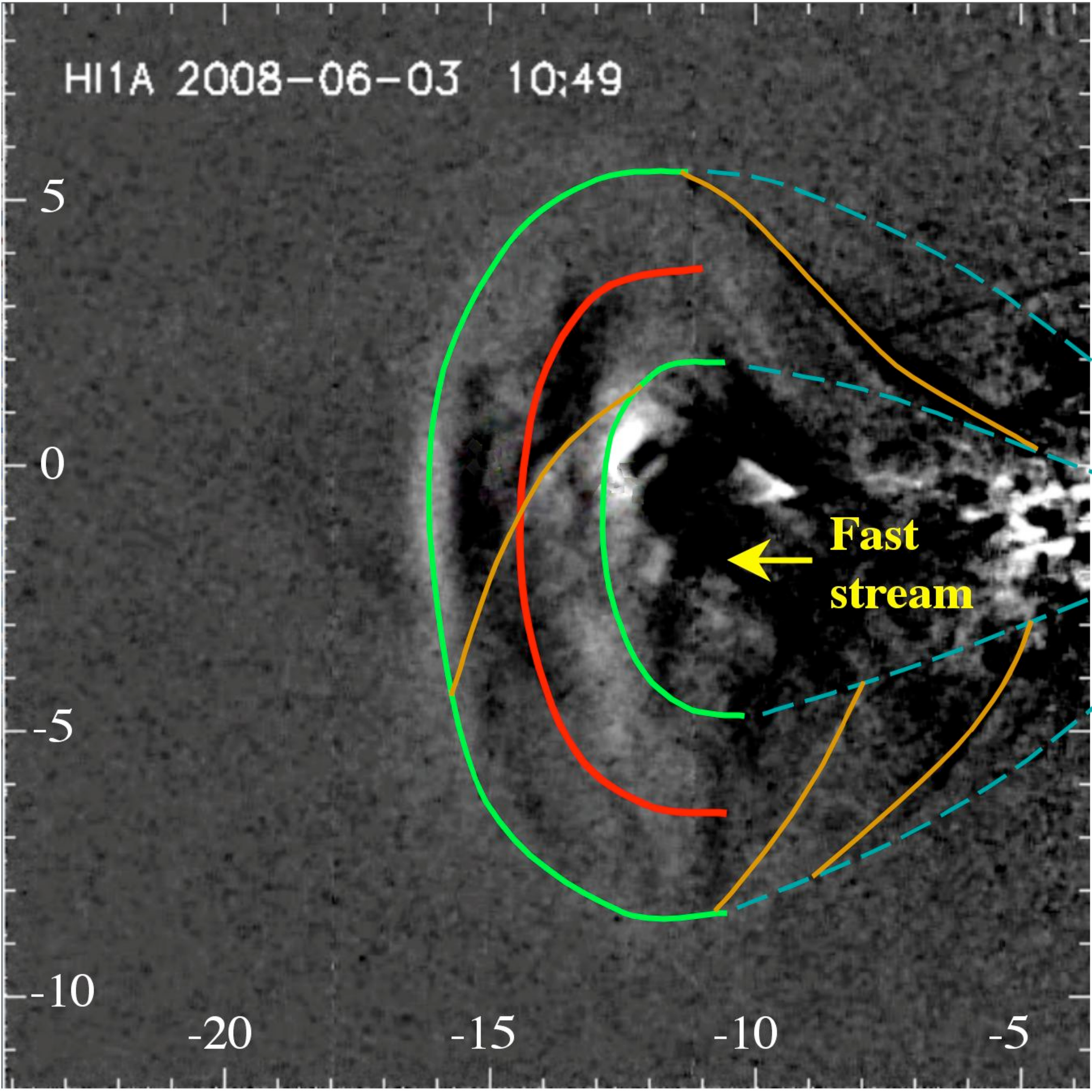}
 \caption{ Observational example of a flux-rope observed by STEREO-A HI1. The image is derived with running differences.  On the right panel, the same image is shown with the front and rear sheaths outlined with green lines (the dashed blue lines are extrapolations towards the Sun).  The flux-rope axis (red line) is defined at equidistance from the two sheaths.  Four twisted-like structures are marked with orange lines.  The coordinate system is the elongation angle in degree from the Sun.  This figure is adapted from \citet{Mostl09c}.}
\label{fig_HI1}
\end{figure*}

\section{Comparison with the results of heliospheric imagers} 
\label{sec_Comparison}

\subsection{Description of the analyzed event} 
\label{sec_C_event}

The heliospheric imagers on board of STEREO provide a 2D view of the strongest density regions.  In the case of CMEs, they typically image the sheath region in front of the CME.  The flux-rope is best seen as an intensity depletion but its extension, even in projection, is typically difficult to define. Indeed, it can for example be partly masked by the sheath and other bright structures present in the background or foreground.  The visualization of the flux-rope requires the development of sophisticated technics to remove the huge background present in the heliospheric images \citep[][and references therein]{Howard12}.

STEREO has so far observed few cases where the flux-rope extension can be estimated.  To our knowledge, the best case for that purpose is the June 2008 event since the flux-rope, observed \insitu\ by STEREO-B, is surrounded by dense plasma which was imaged by STEREO-A from the side (with a longitude difference of $\approx 55^{0}$). {At least two other exceptional cases, 
with density peaks surrounding the flux-rope, have been observed during the two first years of the STEREO mission \citep{Rouillard11b}. However, the June 2008 event remains the most carefully studied amongst those three.}

The associated CME was launched from the Sun on June, 1st 2008 at around 21:00 UT and crossed STEREO-B on June, 6th 2008 at around 23:00 UT, so about 5 days later. It is thus a slow CME \citep{Robbrecht09,Mostl09c} and indeed, the \insitu\ plasma measurements found a mean outward velocity of $\approx 400$ km/s.   A clear rotation of the magnetic field is observed at STEREO-B within a low plasma-$\beta$ region ($\beta \le 0.05$), but this does not strictly define a MC since its proton temperature is comparable to that of the solar wind with similar speed \citep[see Figure 1 of][]{Mostl09c}.  Remarkably, the heliospheric imager of STEREO-A detected density structures which have a twisted appearance all along the flux-rope (see the orange lines in \fig{HI1}).  From their extensions, these dense structures are expected to be at the periphery of the flux-rope. 

The flux-rope is significantly faster than the front solar wind and a shock is present at the front edge of the sheath.   The flux-rope is also overtaken by a faster stream both detected \insitu\ by STEREO-B and the imagers of STEREO-A \citep[see Figure 1 and movies of][]{Mostl09c}.  This fast stream creates a reverse shock, and bound from behind the second sheath region which follows the flux-rope.  Corresponding front and rear sheaths are seen in the heliospheric imagers of STEREO-A as two bright regions bracketing a dim one associated with the flux rope. These imaged sheaths have both two regions of high plasma density as detected \insitu\ by STEREO-B. 
This association has been clearly established by \citet{Mostl09c} by comparing the timing of the \insitu\ enhanced density regions with those of the bright regions when they overtake STEREO-B.
The two external enhanced density regions are expected to be due to plasma compression after the plasma crossed the shocks, while the two internal ones could be due to an earlier over-expansion of the flux-rope.  Then, these two peaks of density in each imaged sheath could be the trace of the propagation and expansion sheaths as defined by \citet{Siscoe08}.

\begin{figure*}
\centering
\sidecaption    
\includegraphics[width=12cm]{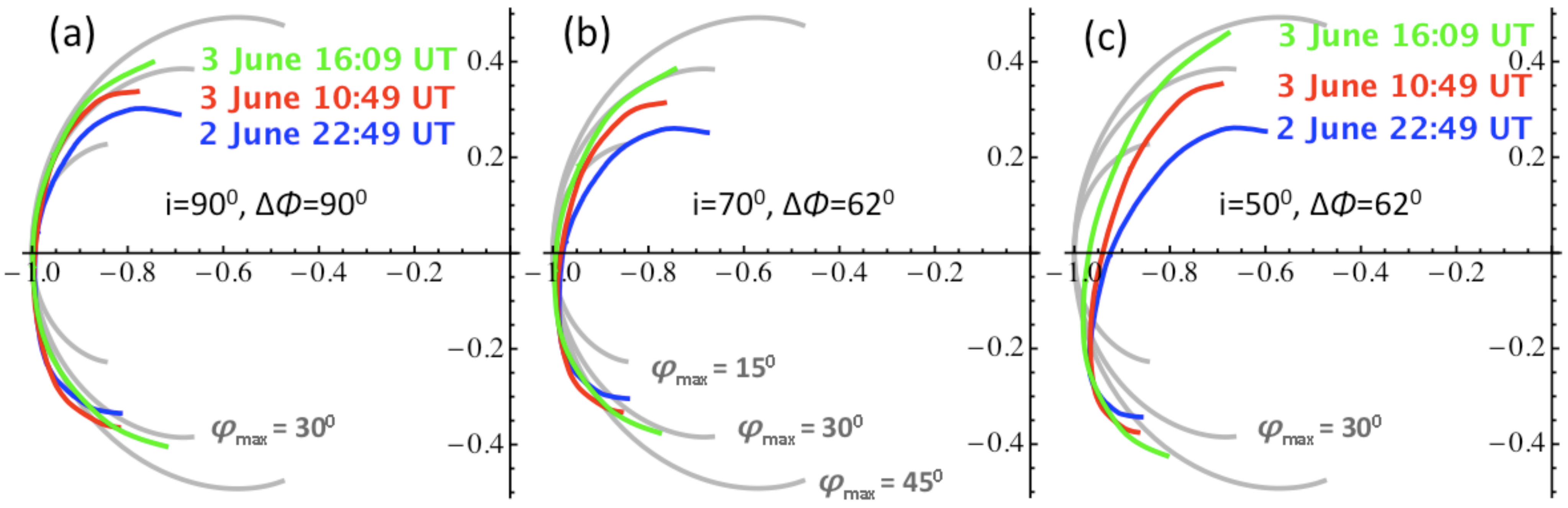}
 \caption{ Comparison of the axis shape deduced from \insitu\ measurements of MCs (grey curves) and from the CME observed by STEREO-A HI1 on 2-3 june 2008 (colored curves, the image corresponding to the red curve is shown in \fig{HI1}).  The panels are in the plane of the flux-rope axis.  We use different geometry to deduce the shape of the axis: (a) conic projection on the plane of the sky, (b-c) axis plane being inclined on the ecliptic by an angle $\iA$ and crossing the ecliptic at a longitude $\Delta \phi$ from STEREO-A (see Sect. \ref{sec_C_compare}).
}
\label{fig_axis_compare}
\end{figure*}

\subsection{Axis shape estimated with imagers} 
\label{sec_C_imagers}

  The relationship found between the in-situ and the imager data implies that the front and rear sheaths bracket the flux rope. In the following, we use this property to estimate the extension of the flux rope from the imagers. 
  
  We manually define the central part of both bright regions on Heliospheric Imager (HI) images and suppose that the flux-rope axis is at mid-distance (\fig{HI1}).  This procedure has large uncertainties.  First, the manual pointing of a bright region has intrinsic bias.  Second, the rear bright region has many structures and is quite difficult to define. Third, the axis may not be exactly at half distance between the two sheaths. Finally, the images are 2D projection of a 3D plasma distribution of unknown shape.   We limit the two first uncertainties as much as possible by repeating independently the pointing on different images taken at different times.
  
  We present results obtained only with HI1 since the contrast of the sheaths with the surrounding regions becomes rapidly faint after the entrance in the HI2 field of view \citep[see the movie attached to][]{Mostl09c}.  Next, the location of the MC axis at about half distance between the sheaths is locally justified by the \insitu\ measurement of the magnetic field and its force-free reconstruction (the flux rope extension is comparable before and after the closest approach to the axis).  Finally, we investigate different geometries to test the projection effect.

The observed bright sheaths are 3D plasma density distribution observed in projection (\fig{HI1}), and we can deduce their curvature by only adding assumptions.  In the line of thinking of \citet{Siscoe08} we consider two extremes. 

One approach is correlated with the hypothesis that the evolution of the sheaths is dominated by the propagation of the ICME. The front one is due to the CME overtaking the slow wind, while the rear sheath {can be} due to the fast wind overtaking the CME. 
Then, we suppose that the two sheaths are part of spherical shells centered on the Sun.   With this simple geometry, the 2D observed shape does not depend on the CME direction and the observed structure is simply a conic projection on the plane of sky of the dense sheaths (\fig{axis_compare}a).

Another approach is to consider the imaged sheaths as a consequence of the flux-rope expansion.
In such a case, a plasma sheath surrounds the flux-rope, and we mostly see the latter when the line of sight is tangent to it.   
In this second approach, we suppose that the observed sheaths are tracing dense plasma located near the plane of the flux-rope.  Both, \insitu\ and imager observations, indicate that STEREO-B crossed the flux-rope close to its apex.  
With a plane inclined on the ecliptic by an angle $\iA$ and which intersection with the ecliptic is at a longitude $\Delta \phi$ from STEREO-A, we project the observations on this plane through a conic projection as viewed from STEREO-A (\fig{axis_compare}b,c).  
In this case an increasing deformation of the flux-rope axis is present as the $\iA$ angle decreases from $90\degree$.

\subsection{Comparison of the axis shape derived from \insitu\ and imager data.} 
\label{sec_C_compare}

The estimations of the axis shape from images obtained above are compared with the results of \sect{Deduction}, and in particular with \fig{axis}. To do so, we have plotted in grey color axis shapes for three $\phimax$ values drawn in the background of \fig{axis_compare}. 
In panel~a, it is remarkable that the simple projection on the plane of sky implies a deduced axis very close to the case $\phimax=30\degree $ at the three times shown (this case is the most suitable when considering the upper/lower (northern/southern) branches of the flux rope axis, with only small deviations in the northern shape at the earliest time).
We have repeated the manual pointing on different images at different observed times.  Even shifting the pointing by a half width of the brightenings, the difference between the deduced shapes is less than between the three southern axis deduced at different times in \fig{axis_compare}a.  

To use the second approach, for which the observed dense plasma is nearby the flux-rope plane, we need to precise the 3D geometry of the event.  STEREO-B observed the flux-rope \insitu\ and its axis was estimated to cut the ecliptic plane slightly eastward, implying that $\Delta \phi \approx 62 \degree$ (with $\Delta \phi$ the longitude angle between the flux-rope axis and STEREO-A).  We find that this angle is sufficiently different from $90 \degree$ to create a significant deformation of the axis when the inclination $\iA$ is significantly different from $90 \degree$.  By fitting the \insitu\ magnetic data with two models, the axis latitude $\theta$ was estimated to be $51\degree$ and $37 \degree$ \citep{Mostl09c}. With a crossing close to the flux-rope apex, the inclination $\iA$ has a comparable value.
Already the case $\iA=70 \degree$ has a marked asymmetry which grows further for $\iA=50 \degree$ (\fig{axis_compare}b,c).  It is unlikely that this asymmetry, present in the flux-rope plane, would be mostly compensated by a projection deformation to provide the nearly symmetric observed shape.  So either the flux-rope axis is more orthogonal to the ecliptic plane than inferred from modeling \insitu\ data (implying a rotation of the whole flux-rope as it evolves from the Sun toward STEREO-B), either STEREO-A observed more two spherical-like dense shells.  
All in all, we find that the error on the definition of the brightening shapes on the images is much smaller than the uncertainty coming from the projection of an unknown 3D shape.

 The \insitu\ proton density measurements of STEREO-B show comparable width and density for the propagation and expansion sheaths, except from a sharp peak, up to twice more dense, for the front expansion sheath \citep[see Fig. 1][]{Mostl09c}.  STEREO-A imagers did not separate the propagation and expansion sheaths both in front and at the rear of the flux-rope, so their interpretation as 3D structures is difficult since two different 3D structures are mixed in one bright structure. Here we suppose that the propagation sheath is part of a spherical shell and the expansion sheath has a tube shaped structure.  While the density is comparable in both, the first one has a larger radius of curvature, so a longer quantity of dense plasma is present along the line of sight.   Then, it is likely that the propagation sheaths are more contributing to the Thomson scattering of the solar light, so that our most relevant axis shape estimation is shown in \fig{axis_compare}a.  The main limitation is that both propagation sheaths do not closely envelope the flux-rope, so the deduced axis shape could have large biases.  Still, the close correspondence found in the axis shape by very different methods is an indication that the systematic bias in each method should not be so large, but they are expected to be of the order of the differences shown in \fig{axis_compare}a.      
     
\section{Summary and Conclusions} 
\label{sec_Conclusion}

  When a spacecraft crosses a MC, detailed \insitu\ measurements of plasma parameters and magnetic field are available only along the spacecraft trajectory.  Global information on the local cross section of the flux-rope are typically derived by solving MHD force-balance equations constrained by the data.  Is it possible to realize a further step to constrain the whole flux-rope? A few multi-spacecraft observations of the same MC have been realized, but they remain case studies and a large number of spacecraft would be required to sample the flux-rope along its axis.  Rather, we have the information on a large set of MCs crossed once at various locations along their flux-ropes.  The present study used this statistical information to derive a mean axis shape (the axis of a particular MC could deviate in respect to this mean shape, but from the method used in the present paper we cannot quantify this deviation).

Our study is based on the results of \citet{Lepping10} and their recent extension to 121 MCs observed by WIND spacecraft over 15 years.  Each MC was fitted with the Lundquist's model.  This fit provides an estimation of the local axis direction (its latitude and longitude). This orientation is an implicit information on the location of the spacecraft crossing along the flux-rope axis.
 In order to precise this, we introduce two new angles to define the local axis direction (\fig{schema}): its inclination on the ecliptic ($\iA$) and its location angle ($\lA$).  If we suppose that the whole axis is planar and loop shaped as in \fig{schema}b,c, then $\iA$ is the inclination of the axis plane on the ecliptic and, going from one leg to the other, $\lA$ evolves continuously from $\approx -90\degree$ to $\approx 90\degree$, with $\lA=0$ at the apex.  Then, $\iA$ and $\lA$ angles are adapted to the geometry of the flux-rope.

Could we analyze together the results of various MCs?  First, we found that the inclination angle ($\iA$) is broadly distributed and we found no significant correlation between $\iA$ and any of the MC parameters.   By contrast the location angle ($\lA$) has a distribution, $\pobs (\lA)$, peaked around zero.  This distribution is almost symmetric, $\pobs (\lA)\approx \pobs (-\lA)$, implying
no significant difference between both legs. We then report results derived from $\pobsl$ in \fig{prob_lambda}.   We further found no significant dependence of $\pobsl$ with $\iA$ angle.  Furthermore, all correlations of MC parameters with $\iA$ angle are very small.  We conclude that the MC properties are independent on the inclination $\iA$ of the flux-rope on the ecliptic.  

The MCs are launched from various solar longitude and moreover the Sun is rotating, so the MC with a low inclination $\iA$ along the ecliptic are expected to be uniformly sampled at random positions by WIND spacecraft (located near Earth).   Describing the supposed planar axis with cylindrical coordinates centered on the Sun ($\rho,\varphi$), it implies that the sampling is expected to be uniform in $\varphi$.  Then, from the observed $\pobsl$ distribution, 
a flux-rope shape can be derived (\sect{Deduction}).   Since $\pobsl$ distribution is not significantly dependent of $\iA$, the hypothesis of uniform angular sampling of a MC set is compatible with the study of the distribution for any sets of MCs considered in this paper (\eg\ \fig{slopePlambda}).   MCs observed with large $\iA$ values are also broadly sampled along their axis because MCs are launched from the Sun from a very broad range of latitude which is mostly kept as the MCs propagate in the interplanetary space (\eg\ Ulysses has observed MCs at latitudes as high as $\approx 80\degree$ in both hemispheres). 

We first test the above idea with a simple global model of the flux-rope axis. Supposing that the axis is part of an ellipse, we found that the distribution of $\lA$ is indeed very sensible to the axis shape.  The observed distribution $\pobsl$ is compatible with an aspect ratio of the ellipse around $1.2$ with the major axis perpendicular to the radial from the Sun, but is incompatible with an aspect ratio of unity (circular shape),
as well as an aspect ratio larger than $1.3$.   In particular the axis is not very flat around its apex (\ie\ $\rho\approx$ constant) since it would imply a distribution $\pobsl$ much more peaked around $\lA=0$ than observed. 

Next, rather than fitting the above specific model of the axis to $\pobsl$, we derive a method to compute the mean shape of the axis directly from $\pobsl$.   Since the shape is derived by integration [\eqs{D_varphi}{D_rho}], the method is robust to any perturbations of $\pobsl$.  Indeed, we verify that very close axis shapes are deduced with various samplings, interpolations and fitting functions of $\pobsl$ distribution (\fig{axis}).  Restricting the MCs set to the best observed ones affect only slightly the axis shape away from the apex.  Our results are compatible with previous results of multi-spacecraft crossings of a MC as summarized in Figs 3 and 4 of \citet{Burlaga90}.
  
It remains one free parameter in the determination of the axis shape from $\pobsl$: the opening angle between the legs ($2~\phimax$, \fig{schema_E}).  The forward modeling with an elliptical shape has shown that the distribution of $\lA$ is weakly affected by $\phimax$. 
 However, the negative side is that $\phimax$ is not constrained by \insitu\ observations, so that $\phimax$ should be provided by another type of observations such as heliospheric imagers.  
  The positive side of this is that sets of MCs with different $\phimax$ can be combined since they have comparable distributions of  $\lA$.  This allows to combine the information of various MCs to build $\pobs (\lA)$ without knowing their $\phimax$ value.  This justifies 
the use of the full set of WIND MCs, or parts of it, to derive a mean axis shape.
 
Heliospheric imagers appear at first better suited to constrain the global shape of MCs.  However,
they image only the dense sheath present in front of the MCs.  In rare cases, a dense sheath is also present at the rear of the MC.  Both sheaths bound the flux-rope allowing in principle to define its global shape. However the 3D shape of the sheath is unknown and they partly overlap the flux-rope along the line of sight, so that a quantitative estimation of the flux-rope shape from imagers needs also hypothesis on the 3D shape of the sheaths.  Moreover, the flux-rope axis is not imaged, then its shape can only be determined indirectly from the sheath locations. 
 
We select a case best suited to define the axis shape from STEREO-A heliospheric imagers, while the flux-rope was also detected \insitu\ by STEREO-B. Supposing that the projection effects are only weakly affecting the observed shape, we derive an axis shape which is comparable to the mean axis shape obtained from $\pobsl$ and $\phimax \approx 30\degree$.
Since these two derivations of the axis shape are based on totally different observing technics, complemented with different hypothesis, the convergence to a comparable axis shape mutually strengthens their results. 
           
The mean axis shape deduced in this work can be used in several applications.   For example,
from the axis orientation, locally determined by modeling, or by fitting a flux rope model to the spacecraft magnetic data, the angle $\lA$ allows to estimate the location of the spacecraft along the  mean axis derived in this study.  This estimates how far from the apex the spacecraft crossing is, in angular distance $\varphi$ (\fig{schema_E}).   Another possible application is the determination of a minimum field line length by linking the ends of the determined axis shape by straight segments connecting to the Sun (other field lines of the flux-rope are longer because of the twist).  This has an application for timing the transport of energetic particles. 
A third direct application is for space weather, as this flux-rope global shape can be incorporated in a kinematic model of CME propagation from the Sun (\eg\ assuming a self similar evolution).  
  
Finally, the method developed in this work could be applied more broadly. We applied it to the results of a Lundquist fit of the \insitu\ data, but it can also be applied to any other method which derives an estimation of the local flux-rope orientation, provided a large enough number of MCs are analyzed. Since the deduced axis shape is mainly determined by the slope of $\pobsl$, which is directly related to the mean of $\lA$, the axis shape estimation does not need a large number of MCs (\eg\ a set of 20 MCs could be sufficient for several applications which require only an approximate shape).
In parallel, it is worth to derive more constraints on the flux-rope shape from imagers, \eg\ by developing the 3D forward models of the flux-rope and its surrounding sheaths.     
   
\begin{acknowledgements}
The work of M.J. is funded by a contract from the AXA Research Fund.
This work was partially supported by the Argentinean grants
UBACyT 20020090100264 and PIP 11220090100825/10 (CONICET) and by a one
month invitation of S.D. by Paris Observatory.
S.D. is member of the Carrera del Investigador Cien\-t\'\i fi\-co, CONICET.
S.D. acknowledges support from the Abdus Salam International Centre
for Theoretical Physics (ICTP), as provided in the frame of his
regular associateship.
\end{acknowledgements}
 
\bibliographystyle{aa}
\bibliography{mc}
\IfFileExists{\jobname.bbl}{}
{\typeout{}
\typeout{****************************************************}
\typeout{****************************************************}
\typeout{** Please run "bibtex \jobname" to optain}
\typeout{** the bibliography and then re-run LaTeX}
\typeout{** twice to fix the references!}
\typeout{****************************************************}
\typeout{****************************************************}
\typeout{}
}

\end{document}